\documentclass[aps,pra,twocolumn,floatfix,superscriptaddress]{revtex4}

\usepackage{amsfonts, amsmath, amssymb}
\usepackage{graphicx}
\usepackage{dcolumn}
\usepackage{bm}
\newcommand{\braket}[1]{\left< #1 \right>}
\newcommand{\bra}[1]{\left< #1 \right|}
\newcommand{\ket}[1]{\left| #1 \right>}
\usepackage[normalem]{ulem}
\usepackage{color}
\usepackage[utf8]{inputenc}
\usepackage[colorlinks,citecolor=blue,linkcolor=blue,urlcolor=blue]{hyperref}

\newcommand{\sket}[1]{|| #1 \rangle\rangle}

\newcommand\red[1]{{\color{red}#1}}

\begin{document}

\title{Cluster mean-field approach to the steady-state phase diagram \\ of dissipative spin systems}

\author{Jiasen Jin}
\affiliation{School of Physics and Optoelectronic Engineering, Dalian University of Technology, 116024 Dalian, China}

\author{Alberto Biella}
\affiliation{NEST, Scuola Normale Superiore and Istituto Nanoscienze-CNR, I-56126 Pisa, Italy}
\affiliation{Kavli Institute for Theoretical Physics, University of California, Santa Barbara, CA 93106, USA}

\author{Oscar Viyuela}
\affiliation{Departamento de F\'isica Te\'orica I, Universidad Complutense, 28040 Madrid, Spain}

\author{Leonardo Mazza}
\affiliation{D\'epartement de Physique, Ecole Normale Sup\'erieure / PSL Research University, CNRS, 24 rue Lhomond, F-75005 Paris, France}
\affiliation{NEST, Scuola Normale Superiore and Istituto Nanoscienze-CNR, I-56126 Pisa, Italy}
\affiliation{Kavli Institute for Theoretical Physics, University of California, Santa Barbara, CA 93106, USA}

\author{Jonathan Keeling}
\affiliation{SUPA, School of Physics and Astronomy, University of St Andrews, St Andrews KY16 9SS UK}
\affiliation{Kavli Institute for Theoretical Physics, University of California, Santa Barbara, CA 93106, USA}

\author{Rosario Fazio}
\affiliation{ICTP, Strada Costiera 11, 34151 Trieste, Italy}
\affiliation{NEST, Scuola Normale Superiore and Istituto Nanoscienze-CNR, I-56126 Pisa, Italy}
\affiliation{Kavli Institute for Theoretical Physics, University of California, Santa Barbara, CA 93106, USA}

\author{Davide Rossini}
\affiliation{NEST, Scuola Normale Superiore and Istituto Nanoscienze-CNR, I-56126 Pisa, Italy}
\affiliation{Kavli Institute for Theoretical Physics, University of California, Santa Barbara, CA 93106, USA}

\begin{abstract}
  We show that short-range correlations have a dramatic impact on the steady-state 
  phase diagram of quantum driven-dissipative systems. 
  This effect, never observed in equilibrium, follows from the fact that ordering 
  in the steady state is of dynamical origin, and is established only at very long time, 
  whereas in thermodynamic equilibrium it arises from the properties of the (free-)energy. 
  To this end, by combining the cluster methods extensively used in equilibrium 
  phase transitions to quantum trajectories and tensor-network techniques, 
  we extend them to non-equilibrium phase transitions in dissipative many-body systems.
  We analyze in detail a model of spins-$1/2$ on a lattice interacting through an XYZ Hamiltonian, 
  each of them coupled to an independent environment which induces incoherent spin flips. 
  In the steady-state phase diagram derived from our cluster approach, the location 
  of the phase boundaries and even its topology radically change, 
  introducing re-entrance of the paramagnetic phase as compared 
  to the single-site mean field where correlations are neglected. 
  Furthermore a stability analysis of the cluster mean-field indicates a susceptibility towards 
  a possible incommensurate ordering, not present if short-range correlations are ignored.  
\end{abstract}


\maketitle

\section{Introduction}
\label{Introduction}

In thermodynamic equilibrium a transition to a state with a spontaneous broken symmetry 
can be induced by a change in the external conditions (such as temperature or pressure) 
or in the control parameters (such as an external applied field). 
The most widely studied examples are for systems at non-zero temperature,
in the framework of classical phase transitions~\cite{goldenfeld}. Here, equilibrium 
thermal fluctuations are responsible for the critical behavior associated with
the discontinuous change of the thermodynamic properties of the system. 
Transitions may also occur at zero temperature, as a function of some 
coupling constant~\cite{sachdev}; in that case, since there are no thermal fluctuations, 
quantum fluctuations play a prominent role. 
For many decades, the study of phase transitions and critical phenomena has attracted 
the attention of a multitude of scientists from the most diverse fields of investigations: 
Phase transitions are present at all energy scales, in cosmology and high-energy 
physics, as well as in condensed matter. 

Moving away from the thermodynamic equilibrium, collective phenomena and ordering also appear 
in open systems, upon tuning the rate of transitions caused by the environment~\cite{weiss}. 
For example, they emerge in most diverse situations~\cite{Cross1993} ranging from the synchronous flashing 
of fireflies~\cite{strogatz} to the evolution of financial markets~\cite{mantegna}. 
The classical statistical mechanics of such driven systems (including traffic models, active matter, 
and flocking) has attracted an increasing attention over the years, see e.g. Refs.~\cite{vicsek,zia}. 
Such interest is in part due to the remarkable possibility of achieving ordered states 
that are not possible in equilibrium systems, displaying for example long-range order 
in two-dimensional flocking~\cite{Toner}, 
something forbidden by the Mermin--Wagner theorem~\cite{Mermin} in equilibrium.

Thanks to the recent impressing experimental progresses, see e.g. Refs.~\cite{kasprzak, baumann, syassen}, 
the investigation of non-equilibrium properties of driven-dissipative systems 
has entered the quantum world. Rydberg atoms in optical lattices~\cite{mueller}, 
systems of trapped ions~\cite{mueller}, exciton-polariton condensates~\cite{Carusotto2013}, 
cold atoms in cavities~\cite{ritsch}, arrays of coupled QED 
cavities~\cite{Houck2012, Tomadin2010-rev}, are probably the most intensively investigated 
experimental platforms in relation to this aim. The predicted steady-state phase diagram 
of these driven dissipative systems becomes incredibly rich, displaying a variety of phenomena. 
Just as for classical statistical mechanics,
phases, which are not possible in an equilibrium phase diagram, may appear~\cite{Lee2013}. 
The steady state itself needs not be time-independent and the system may end up 
in a limit cycle~\cite{Lee2011, Jin2013, Ludwig2013, Chan2015, Schiro2015}. 
Renormalisation group (RG) calculations using the Keldysh formalism have 
been performed~\cite{sieberer}; in some cases the universality class of the transitions 
may be modified both by the presence of the external environment and by non-equilibrium 
effects~\cite{torre, Marino2016}. 
A judicious engineering of the system-bath couplings can lead to 
non-trivial many-body states in the stationary regime~\cite{Diehl2008, Verstraete2009}. 
The field of dissipative many-body open systems embraces a much wider class of problems, 
ranging from transport to relaxation dynamics to quantum information processing 
(just to mention few examples). 
A more comprehensive panorama of the recent literature can be also found in 
Refs.~\cite{Diehl2010, Tomadin2010, Nissen2012, LeBoite2013, Jin2014, Lesanovsky2010, hartmann2010, 
Grujic2012, carusotto2009, Umucalilar, Daley2014, Hoening2014, Hoening2013, Petrosyan2013, 
Biella2015, Maghrebi2015, Ilievski2014, Prosen2014, Sibalic2015, Hafezi2016} and citations therein.

In condensed matter systems, most notably in Josephson junction arrays, 
the impact of an external bath on the phase diagram, and the relative critical properties 
was thoroughly studied over the last twenty years, 
see e.g.~\cite{Chakravarty1986, Korshunov1989, Bobbert1992, Refael2007}. 
In all those studies the system and the bath were in an overall equilibrium situation 
at a given (possibly zero) temperature. In quantum driven-dissipative 
systems, such as that considered here, non-equilibrium conditions 
and the flow of energy through the system play a major role.

Our work focuses on an important aspect of the physics of many-body open systems: 
the determination of the steady-state phase diagram. We are going to consider systems 
in which the coupling to the environment leads to a Markovian dynamics. In these cases 
the evolution of the corresponding density matrix $\rho(t)$ obeys the Lindblad equation
\begin{equation}
  \frac{\partial \rho}{\partial t} = -\frac{i}{\hbar} [\hat H,\rho] 
  + \sum_j{\cal L}_j[\rho] \,.
  \label{eq:Master_Eq}
\end{equation}
The first term in the r.h.s. describes the coherent unitary time evolution 
(ruled by the system Hamiltonian $\hat H$). 
The second term, corresponding to a sum of Lindbladian superoperators ${\cal L}_j[\rho]$, 
takes into account the coupling to the external bath(s). The steady-state phase diagram 
is obtained by looking at the long-time limit ($t \rightarrow \infty$) 
of the solution to Eq.~\eqref{eq:Master_Eq} and computing appropriate averages 
$\langle {\hat O} \rangle = \mbox{Tr} \big[ {\hat O} \, \rho_{t \rightarrow \infty} \big] \equiv \mbox{Tr} \big[ {\hat O} \, 
\rho_{\scriptscriptstyle \rm SS} \big]$ of local observables ${\hat O}$, in order to determine 
the (possible) existence of phases with broken symmetries (space, time, spin, \ldots)~\cite{topological}. 

Nearly all the results obtained so far on the phase diagram 
(with the notable exception of the works based on Keldysh RG mentioned above) 
rely on the (single-site) mean-field approximation, where all the correlations are ignored. 
Very little is known beyond that limit about the interplay between many-body correlations and dissipation, 
although there are some contributions in this direction~\cite{Finazzi2015, Degenfeld2014, Weimer2015}.
While quasi-exact numerical methods exist for open one-dimensional systems, 
unfortunately no true phase transitions are expected to occur in that context. 
Beyond one dimension, such methods are much harder to apply. 
However it is well known that the mean-field decoupling, while important to grasp 
the salient features of the system, is not at all accurate in locating the phase boundaries. 

An improvement in the determination of the phase diagram can be obtained by 
a systematic inclusion of short-range correlations (up to a given cluster size). 
In equilibrium, this has been achieved within the cluster mean-field 
approximation~\cite{Bethe1935, Kikuchi1951, Oguchi1955}, 
and using linked cluster expansions~\cite{Oitmaa2006}. 
In the cluster mean-field approach, the accuracy of the diagram is obviously related 
to the size of the considered cluster. Even though it is still mean-field in nature, 
a suitable scheme that combines it with finite-size scaling may in principle allow 
to extract non-classical critical exponents~\cite{Suzuki1986}. 
In higher dimensions (above the lower critical one) where one expects spontaneous 
symmetry breaking, cluster methods lead only to quantitative corrections (a mere shift) 
to the mean-field predictions. 
These corrections become smaller on increasing the dimensionality. 

For equilibrium phase transitions, the topology of the phase diagram 
is well captured at the mean-field level, and the short-range fluctuations considered 
by cluster methods only lead to shifts in the location of the transition lines/points. 
Normally, they do not cut an ordered phase into two separate parts, divided by a disordered region. 
The possibility to have a radical change of topology is however permitted out of equilibrium,
where the spontaneous breaking of symmetry is of pure dynamical nature: 
terms which are formally irrelevant in the RG sense can nonetheless modify the flow 
of RG-relevant terms, so as to move a point in parameter space from one side 
to the other side of a phase boundary. 
Such a scenario is rarely, if ever, seen in equilibrium.

\begin{figure}[!t]
  \includegraphics[width=0.98\columnwidth]{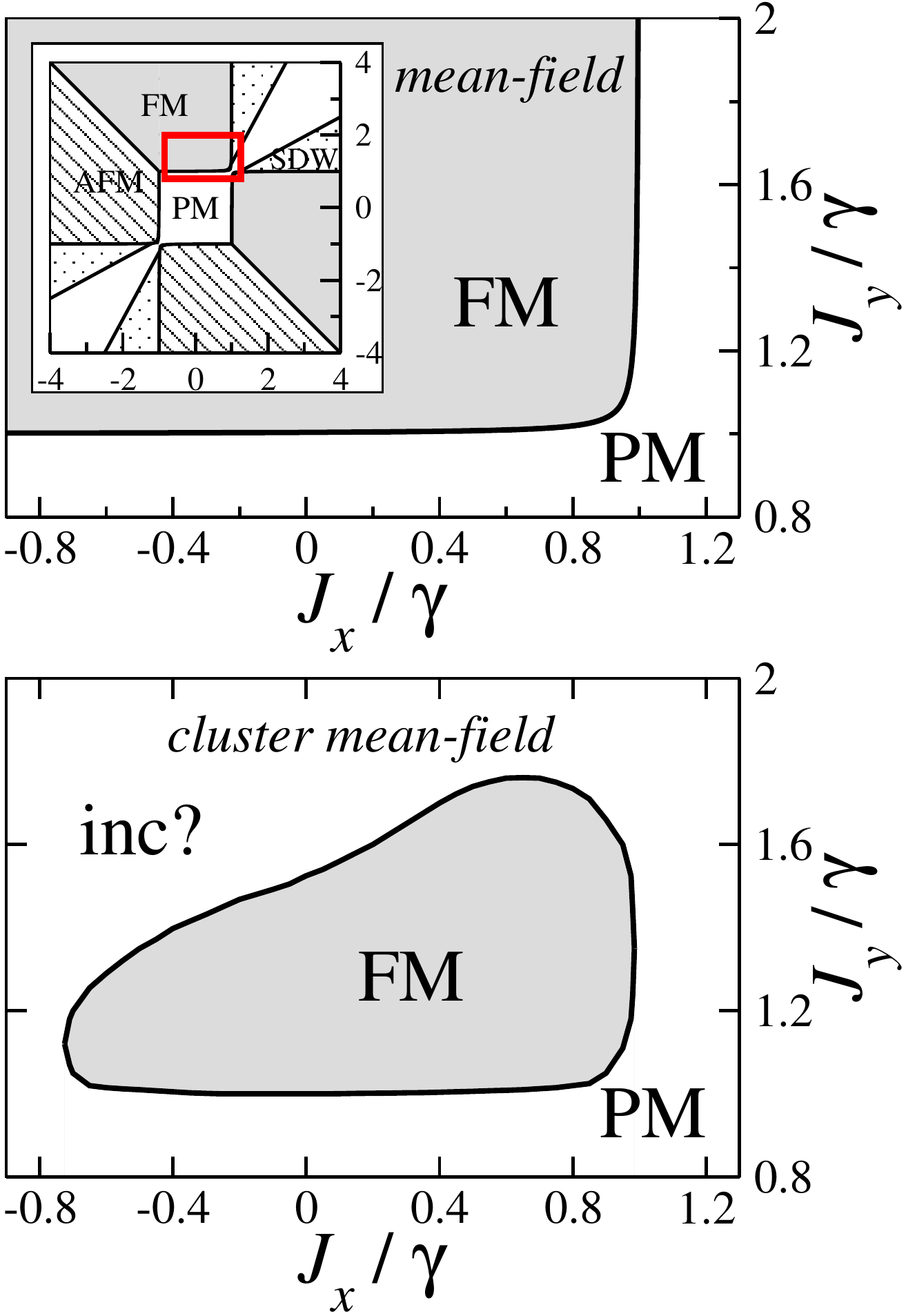}
  \caption{A sketch of the phase diagram of the model defined by Eqs.~\eqref{eq:HamSpin}
    and~\eqref{eq:Lindblad}, for $J_z=1$. The single-site mean-field as worked out 
    in Ref.~\cite{Lee2013} would predict the emergence of different phases: 
    paramagnetic (PM), ferromagnetic (FM), antiferromagnetic (AFM), and spin-density-wave (SDW)
    (inset to the top panel).
    Here we focus on the region highlighted by the red box, which displays
    a transition from PM to FM states (magnified in the top panel).
    A proper inclusion of short-range correlations (through the cluster mean-field) 
    shrinks the ferromagnetic region to a small ``island'', thus suppressing the order 
    at large couplings and hinting to a possible incommensurate (inc) ordering
    (bottom panel).}
  \label{fig:diagram}
\end{figure}

We demonstrate that the above picture is indeed verified in the open many-body context,
and ordering with a non-trivial spatial pattern may emerge (see Fig.~\ref{fig:diagram}). 
The most natural way to show this is to include correlations through a cluster mean-field
analysis which, to the best of our knowledge, has never been systematically applied
in the open many-body context. 
Although the general strategy is the same as for equilibrium systems, there are several 
peculiarities emerging in this scenario, which need to be carefully addressed. 
The steady-state solution typically needs to be obtained dynamically via 
Eq.~\eqref{eq:Master_Eq} (and not through a solution of a self-consistent equation~\cite{note01}). 
To increase the cluster size, we introduce a new approach that combines 
the cluster mean field with quantum trajectories~\cite{Dalibard1992} 
and with matrix-product-operators~\cite{Verstraete2004, Zwolak2004}.

We apply our technique to a spin-1/2 XYZ-model with relaxation (as previously studied 
by Lee {\it et al.}~\cite{Lee2013}), and show that the short-range correlations captured 
by the cluster approach can have a dramatic effect on the phase diagram. 
This last point is exemplified in Fig.~\ref{fig:diagram} (that summarises one of our main results).
A mean-field analysis predicts a transition from a paramagnet to a ferromagnet (upper panel) 
in the whole region of large couplings $J_y > J_y^{c}$. The lower panel sketches the outcome 
of the cluster analysis. The ferromagnetic regime has shrunk to a finite region 
disappearing in the limit of large couplings.
For an equilibrium system, such behavior would be very strange: large coupling strengths 
increase the tendency toward ferromagnetic order, yet here we find that the ordered state 
is destroyed by strong couplings.
Furthermore, indications from a stability analysis hint at a different type of ordering
at large values of $J_y$.

The paper is organized as follows. In the next section we define the spin-1/2 model with 
nearest-neighbor XYZ interactions coupled to a local bath, which will be considered 
in the following.
We then introduce the cluster mean-field approach to driven dissipative systems and show 
how to combine it with quantum trajectories (Sec.~\ref{subsec:qt}) 
and with the matrix-product-operator (Sec.~\ref{subsec:mpo}) formalism. 
We will see this method at work by looking at the steady-state phase diagram 
and comparing its rich features with those pointed
out in Ref.~\cite{Lee2013} at the single-site mean-field level. 
Specifically, in Sec.~\ref{results} we discuss how the location of the transition lines 
is qualitatively changed in the cluster approach. Our aim is to highlight the key role 
of short-range correlations in driven-dissipative systems. For this purpose,
we will concentrate on a specific region of the diagram where a paramagnetic to 
ferromagnetic transition takes place. 
In one dimension (Sec.~\ref{sec:1D}) the cluster approach with appropriate scaling 
restores the absence of symmetry breaking. 
While the one-dimensional results presented here are as could be expected, 
we believe they are however useful as a benchmark of the numerical methods 
employed in the rest of this paper.
Surprises appear in the 
two-dimensional case (Sec.~\ref{sec:2D}), where a ferromagnetic phase is possible. 
Including cluster correlations gives rise to a phase diagram radically different 
from what derived within single-site mean-field. 
The extent of the ferromagnetic region becomes finite.
The nature of the such transitions is discussed in Sec.~\ref{sec:stability}, 
where a stability analysis around the mean-field solution is performed. 
The finite extent of the ordered phase appears to
persist in higher-dimensional systems (Sec.~\ref{sec:HigherD}), 
even though the mean field progressively becomes, as expected, more accurate. 
The underlying dynamical mechanism responsible for such dramatic modifications 
in the phase diagram will be discussed in Sec.~\ref{sec:physicalorig} where we will provide 
a more physical intuition of the results obtained in this work.
Finally in Sec.~\ref{sec:concl} we conclude with a brief summary of our results.

\section{The Model} 
\label{sec:Model}

We consider a spin-$1/2$ lattice system whose coherent internal dynamics is governed by 
an anisotropic XYZ-Heisenberg Hamiltonian,
\begin{equation}
  \hat H = \sum_{\langle i,j\rangle} h_{ij} 
  = \sum_{\langle i,j\rangle} \left( J_x \hat \sigma_i^x \hat \sigma_j^x 
  + J_y \hat \sigma_i^y \hat \sigma_j^y + J_z \hat \sigma_i^z \hat \sigma_j^z \right) \,,
  \label{eq:HamSpin}
\end{equation}
$\hat \sigma_j^\alpha$ ($\alpha = x,y,z$) denoting the Pauli matrices on the $j$-th site 
of the system. The Lindbladian for this model reads
\begin{equation}
\sum_j {\cal L}_j [\rho] 
  = \gamma \sum_{j} \bigg[ \hat \sigma_j^- \rho \, \hat \sigma_j^+
    - \frac{1}{2} \left\{ \hat \sigma_j^+ \hat \sigma_j^- , \rho \right\} \bigg] \,,
  \label{eq:Lindblad}
\end{equation}
where $\gamma$ is the rate of the dissipative processes that tend to flip all the spins down 
independently [$\hat \sigma_j^\pm = \frac{1}{2} \left( \hat \sigma_j^x \pm \hat \sigma_j^y \right)$
stand for the corresponding raising and lowering operators along the $z$ axis]. 
In the rest of the paper we set $\hbar = 1$ and work in units of $\gamma$. 
The (single-site) mean-field phase diagram of the model defined in Eqs.~\eqref{eq:HamSpin} 
and~\eqref{eq:Lindblad} has been worked out in Ref.~\cite{Lee2013}; 
for orientation we summarise the main results of this analysis here.

It is important to remark that an in-plane XY anisotropy ($J_x \neq J_y$) is fundamental 
to counteract the dissipative spin flips along the orthogonal direction~\cite{Lee2013}. 
In the case in which $J_x = J_y$, Eq.~\eqref{eq:HamSpin} reduces to an XXZ Heisenberg model. 
Since this latter conserves the global magnetisation along the $z$ axis, the steady-state 
solution $\rho_{\scriptscriptstyle \rm SS}$ of Eq.~\eqref{eq:Master_Eq} would trivially coincide 
with the pure product state having all the spins aligned and pointing down along the $z$ direction.
This corresponds to a paramagnetic state where the dissipation is dominant, and such that 
$\langle \hat \sigma^x_j \rangle_{\scriptscriptstyle \rm SS} = \langle \hat \sigma^y_j \rangle_{\scriptscriptstyle \rm SS} = 0$
and $\langle \hat \sigma^z_j \rangle_{\scriptscriptstyle \rm SS} = -1$, where 
$\langle {\hat O} \rangle_{\scriptscriptstyle \rm SS} = {\rm Tr} \big( {\hat O} \rho_{\scriptscriptstyle \rm SS} \big)$ 
denotes the expectation value of a given observable ${\hat O}$ on the steady state.

The steady-state phase diagram presented in Ref.~\cite{Lee2013}
is particularly rich and includes, for strongly anisotropic spin-spin interactions, ferromagnetic, 
antiferromagnetic, spin-density-wave, and staggered-XY states. 
Hereafter we concentrate in the regime of parameters $J_x , J_y \ge 1$ and $J_z=1$, where the 
single-site mean field predicts a single ferromagnetic (FM) to paramagnetic (PM) phase transition. 
Indeed by changing the various coupling constants, the PM phase may become unstable 
and the system can acquire a finite magnetisation along the $xy$ plane 
($\langle \hat \sigma^x_j \rangle_{\scriptscriptstyle \rm SS}, \, \langle \hat \sigma^y_j 
\rangle_{\scriptscriptstyle \rm SS} \neq 0$), thus entering a FM phase. 
This fact is associated to the spontaneous breaking of the ${\mathbb Z}_2$ symmetry 
which is present in the model, and corresponds to a $\pi$ rotation along the $z$ axis 
($\hat \sigma^x \to - \hat \sigma^x$, $\hat \sigma^y \to - \hat \sigma^y$). 
The picture changes dramatically when local correlations are included. 

As already mentioned in the introduction, 
in an open system the stationary state may also break time-translational invariance 
(the steady state is time periodic)~\cite{Lee2011, Jin2013, Ludwig2013, Chan2015, Schiro2015}. 
Our numerics suggests that a time-independent solution exists for all parameters we study, and so
we will not consider this last case and concentrate on stationary time-independent solutions. 
This corresponds to the stationary point of Eq.~\eqref{eq:Master_Eq}, 
$\partial_t \rho_{\scriptscriptstyle \rm SS} = 0$, irrespective of the initial condition. 
In the remainder of the paper, we will always implicitly refer to this occurrence.

\section{Methods} 
\label{sec:methods}

Solving Eq.~\eqref{eq:Master_Eq} for a many-body system is a formidable task, 
even from a numerical point of view. 
The exponential increase of the Hilbert space makes a direct integration 
of the master equation unfeasible already for relatively small system sizes. 
Indeed one needs to manipulate a density matrix of dimensions $2^L \times 2^L$, which becomes 
a computationally intractable task already for quite small number of sites ($L \gtrsim 10$). 
In order to access systems as large as possible and to perform finite-size scaling 
up to reasonable sizes, we employ a combination of strategies.

In this section we discuss how to use cluster mean-field methods for driven-dissipative systems; 
these will be employed to determine the phase diagram of the model defined 
by Eqs.~\eqref{eq:HamSpin} and~\eqref{eq:Lindblad}. 
In order to keep the notation as simple as possible, we will describe the cluster approach 
in the spin-$1/2$ language for nearest neighbor Hamiltonians. A straightforward extension 
of our formalism allows to consider generic short-range Hamiltonians of the form 
$\hat H = \sum_{i} \hat h^{(0)}_i + \sum_{\langle i,j \rangle} \hat h^{(1)}_{ij} 
+ \sum_{\langle \langle i,j \rangle \rangle} \hat h^{(2)}_{ij} + \dots$ 
(with the various terms respectively including on-site, nearest-neighbor, 
next nearest-neighbor, \ldots, couplings) and a generic dissipator containing more than 
one Lindblad operator on each site. 

\subsection{Cluster mean-field}
\label{sec:mf}

\begin{figure}[!t]
  \includegraphics[width=0.95\columnwidth]{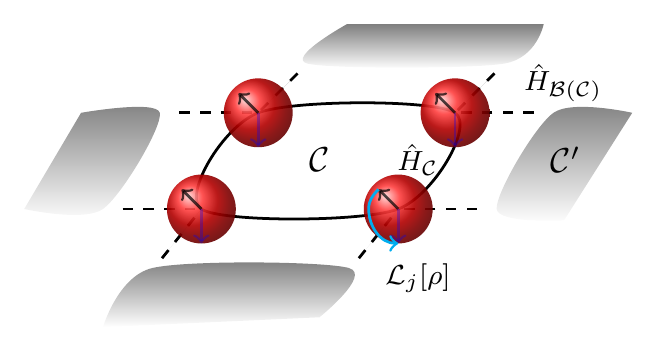}
  \caption{Sketch of the cluster mean-field approach in a dissipative system 
    of interacting spin-1/2 particles. 
    The figure refers to $2 \times 2$ cluster on a two-dimensional square lattice.}
  \label{fig:CMF-sketch}
\end{figure}

Let us isolate a given subset ${\cal C}$ of contiguous lattice sites, 
hereafter called {\it cluster}, from the rest of the lattice forming the system
(which is supposed to be at the thermodynamic limit). 
This is pictorially shown in Fig.~\ref{fig:CMF-sketch}.
The decoupled cluster mean-field (CMF) Hamiltonian with respect to the cluster 
can be written as 
\begin{equation}
  \hat H_{\scriptscriptstyle \rm CMF} = \hat H_{\scriptscriptstyle \cal C} + \hat H_{\scriptscriptstyle {\cal B}({\cal C})} \,,
  \label{eq:H_cmf}
\end{equation}
where
\begin{equation}
  \hat H_{\scriptscriptstyle \cal C} = \sum_{\langle i,j \rangle \vert i,j \in \cal C} \hat h_{ij} 
  \label{eq:Ham_cl}
\end{equation}
faithfully describes the interactions inside the cluster, while
\begin{equation}
  \hat H_{\scriptscriptstyle {\cal B}({\cal C})} 
  = \! \sum_{ j \in {\cal B}({\cal C})} \! {\bf B}^{\rm eff}_j \cdot \hat {\boldsymbol \sigma}_j
\label{eq:CMF}
\end{equation}
effectively represents the mean-field interactions of the cluster ${\cal C}$ with its neighbors
[$\hat {\boldsymbol \sigma}_j = (\hat \sigma^x_j, \hat \sigma^y_j, \hat \sigma^z_j)$]. 
The sum is restricted to the sites on the boundary ${\cal B}({\cal C})$ of the cluster. 
The parameter ${\bf B}^{\rm eff}_j = (B^x_j, B^y_j, B^z_j)$ in Eq.~\eqref{eq:CMF} 
is related to the average magnetisation of the neighboring spins of $i$ belonging 
to the cluster ${\cal C'}$ adjacent to ${\cal C}$. The effective field 
needs to be computed self-consistently in time. 

This reduced description 
arises from a factorized Ansatz for the global density matrix 
\begin{equation}
  \rho_{\scriptscriptstyle \rm CMF} = \bigotimes_{\cal C} \rho_{\scriptscriptstyle \cal C} \,,
  \label{eq:rho_cmf}
\end{equation}
where $\rho_{\scriptscriptstyle \cal C}$ is the density matrix of the ${\cal C}$-th cluster.
Inserting such Ansatz into Eq.~\eqref{eq:Master_Eq} and exploiting 
the translational invariance with respect to the cluster periodicity 
($\rho_{\scriptscriptstyle \cal C}=\rho_{\scriptscriptstyle \cal C'}, \ \forall \ {\cal C,C'}$) 
we get an effective master equation of the form: 
\begin{equation}
  \frac{\partial \rho_{\scriptscriptstyle \cal C}}{\partial t} = -\frac{i}{\hbar} [\hat H_{\scriptscriptstyle \rm CMF},\rho_{\scriptscriptstyle \cal C}] 
  + \sum_{j \in \cal C} {\cal L}_j[\rho_{\scriptscriptstyle \cal C}] \,.
  \label{eq:Master_cluster}
\end{equation}
We recall that the standard mean-field treatment derives from assuming that the cluster is formed by a single site. 

The mean-field approach represents a crude approximation for a many-body interacting system, 
since all the correlations are effectively neglected. The decoupling on a larger structure 
described above partially overcomes this problem: the idea is that interactions 
among the sites inside a cluster are treated exactly [see Eq.~\eqref{eq:Ham_cl}], 
while those among neighboring clusters are treated 
at the mean-field level [see Eq.~\eqref{eq:CMF}].
As a consequence, short-range correlations inside the cluster are safely taken into account. 
The full problem is eventually simplified into the evolution of the 
density matrix $\rho_{\scriptscriptstyle \cal C}$ of the cluster in the presence 
of a time-dependent effective field ${\bf B}^{\rm eff}_j(t)$. 

So far what we discussed equally applies to any cluster mean-field approximation, 
either classical or quantum. The only non trivial modification in the present case 
is that one has to study the evolution of Eq.~\eqref{eq:Master_cluster} 
in the presence of a time-dependent field that has to be determined self-consistently. 
In order to improve its accuracy and to have a reliable scaling of the correlations, 
clusters of sufficiently large dimensions need to be considered. 
For small clusters a direct integration of the cluster master equation is feasible, 
while larger clusters can be faithfully treated by combining the above explained approach 
with specific techniques designed to deal with open systems.
Specifically, we are going to integrate the cluster mean-field approximation 
together with quantum trajectories and with tensor-network approaches. 
The idea and procedure is straightforward, but 
some practical details require stating explicitly.
We present such details in the next sections.

\subsection{Quantum trajectories}
\label{subsec:qt}

There is a simple procedure that allows to avoid simulating the mixed time evolution 
of the full master equation~\eqref{eq:Master_Eq} [which would need to store and evolve 
a $2^L \times 2^L$ matrix $\rho(t)$]. Indeed it can be shown that one can equivalently 
perform a stochastic evolution protocol of a pure state vector of size $2^L$, 
according to the quantum-trajectory (QT) approach~\cite{Dalibard1992}
[which requires to manipulate $N \times 2^L$ elements, $N$ being the number of trajectories
(typically $N \ll 2^L$ is sufficient to get reliable results)]. 
The unitary time evolution part of Eq.~\eqref{eq:Master_Eq}, 
together with the anti-commutator term in Eq.~\eqref{eq:Lindblad}, can be regarded as if 
the evolution were performed by means of an effective non-Hermitian Hamiltonian 
$\hat H_{\rm eff} = \hat H + i \hat{K}$, with 
$\hat{K} = - \frac{\gamma}{2} \sum_j \hat{\sigma}^+_j \hat{\sigma}^-_j$. 
The remaining term in Eq.~\eqref{eq:Lindblad} gives rise to the so-called quantum jumps.
It can be shown that, if the density matrix at some reference time $t_0$ is given by 
the pure state $\rho(t_0) = \ket{\psi_0} \bra{\psi_0}$, after an infinitesimal 
amount of time $\delta t$ it will evolve into the statistical mixture of the pure states 
$\{ \ket{\tilde \psi_0}, \ket{\tilde \psi_j} \}_{\scriptscriptstyle j = 1, \ldots, L}$ 
(the tilde indicates states at time $t_0 + \delta t$):
\begin{equation}
  \rho(t_0 + \delta t) \! = \! \Big( 1 - \! \sum_j {\rm d} p_j \Big) \ket{\tilde \psi_0} \bra{\tilde \psi_0}
  + \! \sum_j {\rm d} p_j \ket{\tilde \psi_j} \bra{\tilde \psi_j} \, ,
  \label{eq:rhot_QT}
\end{equation}
where ${\rm d} p_j = \red{\gamma} \bra{\psi_0} \hat{\sigma}^+_j \hat{\sigma}^-_j \ket{\psi_0}$ and
\begin{equation}
  \ket{\tilde \psi_0} = \frac{e^{-i \hat H_{\rm eff} \delta t}\ket{\psi_0}}{\sqrt{1 - \sum_j {\rm d} p_j}} \,, \quad
  \ket{\tilde \psi_j} = \frac{\hat{\sigma}^-_j \ket{\psi_0}}{\lVert \hat{\sigma}^-_j \ket{\psi_0} \lVert} \,.
\label{eq:Psi_QT}
\end{equation}
Therefore, with probability ${\rm d}p_j$ a jump to the state $\ket{\tilde \psi_j}$ occurs, 
while with probability $1 - \sum_j {\rm d} p_j$ there are no jumps and the system evolves 
according to $\hat H_{\rm eff}$. Assuming that there exists a single steady state 
$\rho_{\scriptscriptstyle \rm SS}$ for Eq.~\eqref{eq:Master_Eq}, one has~\cite{Dalibard1992}:
\begin{equation}
  {\rm Tr} (\hat{O} \rho_{\scriptscriptstyle \rm SS}) =
  \lim_{T \to \infty} \frac{1}{T} \int_{T_0}^{T_0+T} \bra{\psi(t)} \hat{O} \ket{\psi(t)} {\rm d}t\,,
\end{equation}
for any observable $\hat{O}$ and \red{reference time $T_0$}. 
The state $\ket{\psi(t)}$ is stochastically chosen among those in Eq.~\eqref{eq:Psi_QT}, 
according to the statistical mixture~\eqref{eq:rhot_QT}, after iterating the above algorithm 
for $(t-t_0)/\delta t$ times, where the time interval $\delta t$ has to be much smaller 
than the relevant dynamical time scales.

\begin{figure}[!t]
  \includegraphics[width=0.95\columnwidth]{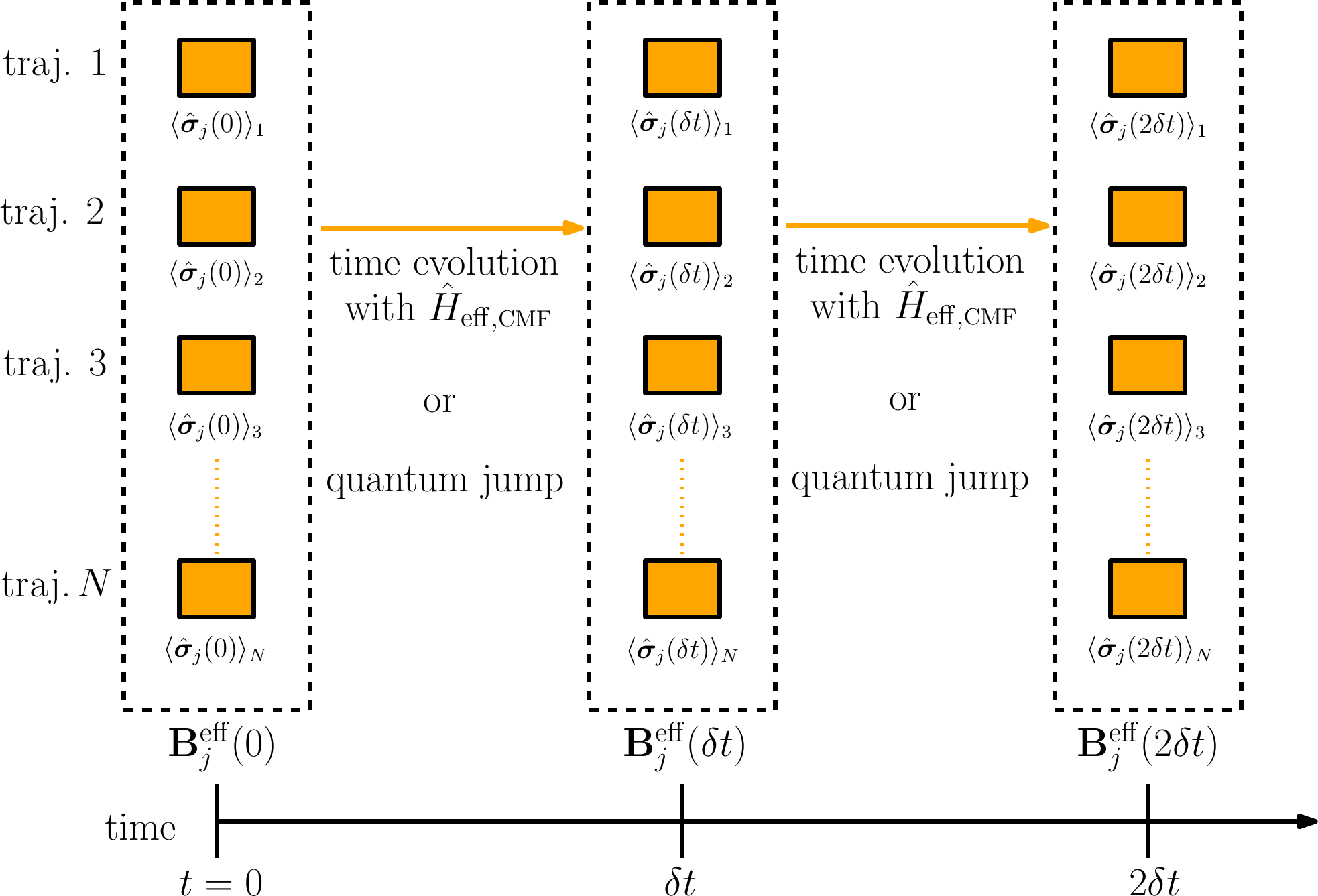}
  \caption{Quantum trajectories combined with mean-field / cluster mean-field method. 
    Coloured boxes along a given line stand for the time-evolved state of the $k$-th trajectory,
    which is stochastically chosen among the set of pure states 
    $\{ \ket{\psi_0(t)}_k, \ket{\psi_j(t)}_k \}$ according to Eq.~\eqref{eq:rhot_QT}. 
    For each of those states, one finds the corresponding mean fields on each site $j$ 
    inside the considered cluster, $\langle \hat{\boldsymbol \sigma}_j(t) \rangle_k$.
    The mean field ${\bf B}^{\rm eff}_j(t)$ parametrizing the effective Hamiltonian
    in Eq.~\eqref{eq:Heff_QT-CMF}, to be used in the master equation for the cluster
    density matrix, is obtained by averaging over all the $N$ trajectories.}
  \label{fig:QT-CMF}
\end{figure}

It is possible to combine the QT method with the above described CMF approach 
at the cost of some moderate modifications. In order to do that, it is necessary 
to perform a simulation of a sufficiently large number $N$ of trajectories in parallel. 
For each trajectory $k$, the mean-field expectation value 
$\langle \hat{\boldsymbol \sigma}_j(t) \rangle_k \equiv {}_k \bra{\psi(t)} \hat{\boldsymbol \sigma}_j \ket{\psi(t)}_k$ 
on each site $j$ of the considered cluster ${\cal C}$ is computed iteratively in time. 
The average over all the trajectories gives the correct mean field at time $t$,
\begin{equation}
  {\bf B}^{\rm eff}_j(t) = \frac{1}{N} \sum_{k=1}^{N} \langle \hat{\boldsymbol \sigma}_j(t) \rangle_k \,.
\end{equation}
which has to be self-consistently used to describe effective interactions 
between adjacent clusters [see Eq.~\eqref{eq:CMF}].
Note that this approach corresponds to performing the stochastic
unravelling of the cluster mean field theory. Such an approach is different from
performing a cluster mean-field decoupling of a stochastic unravelling
of the original equation (i.e., each trajectory would evolve according to {\it its own} mean field). 

Eventually one gets an effective non-Hermitian cluster mean-field Hamiltonian
\begin{equation}
  \hat H_{\rm eff, \scriptscriptstyle CMF} = \big( \hat H_{\scriptscriptstyle \cal C} 
  + i \hat{K}_{\scriptscriptstyle \cal C} \big) + 
  \hat H_{\scriptscriptstyle \cal B(C)} \,,
  \label{eq:Heff_QT-CMF}
\end{equation}
which, together with the possibility of having quantum jumps, governs the time evolution 
of each trajectory for the next time step, as in Eqs.~\eqref{eq:rhot_QT}-\eqref{eq:Psi_QT}. 
The idea of this combined approach is schematically depicted in Fig.~\ref{fig:QT-CMF},
and turns out to be effective to deal with clusters containing $L \gtrsim 10$ sites.

\subsection{Matrix product operators}
\label{subsec:mpo}

Quantum trajectories are not the only method which can be fruitfully combined 
to cluster mean-field techniques. Tensor networks are also ideally suited to this aim.
Below we will consider Matrix Product Operators (MPO) that work very well 
for one-dimensional (1D) systems. 
It would be highly desirable to have tensor-network approaches also in higher dimensions. 
We believe that in combination with cluster mean-field, this will represent a significant 
step forward in an accurate analysis of this class of non-equilibrium critical points.

For 1D systems, the long-time limit of Eq.~\eqref{eq:Master_Eq} can be faithfully addressed 
using a MPO Ansatz for the density matrix~\cite{Verstraete2004, Zwolak2004}. 
The solution $\rho_{\scriptscriptstyle \rm SS}$ is reached dynamically by following 
the time evolution according to Eq.~\eqref{eq:Master_Eq}, using an algorithm based 
on the time-evolving block decimation (TEBD) scheme~\cite{VidalTEBD} adapted to open systems.

The starting point is based on the fact that a generic many-body mixed state on a $L$-site lattice, 
$\rho = \sum_{\vec i, \vec j} C_{i_1 \cdots i_L, \, j_1 \cdots j_L} \ket{i_1 \cdots i_L} \bra{j_1 \cdots j_L}$ 
(we defined $\vec i = \{ i_1 \ldots i_L \}$) can be written as a matrix product state 
in the enlarged Hilbert space of dimension $d^L \otimes d^L$, where $d$ is the dimension 
of the onsite Hilbert space.
By means of repeated singular value decompositions of the tensor $C_{i_1 \cdots i_L, \, j_1 \cdots j_L}$, 
it is possible to obtain
\begin{eqnarray}
  \rho_{\rm \scriptscriptstyle MPO} \!\!\! & = \!\!\!\! & \sum_{\vec i, \vec j = 1}^d 
  \sum_{\vec \alpha=1}^{\chi} \ (\Gamma_{1,\alpha_1}^{[1]i_1,j_1} \, \lambda_{\alpha_1}^{[1]}) 
  \, (\Gamma_{\alpha_1,\alpha_2}^{[2]i_2,j_2} \, \lambda_{\alpha_2}^{[2]}) \dots \cr
  &&\cr
  &&\dots (\lambda_{\alpha_{L-1}}^{[L-1]}\Gamma_{\alpha_{L-1},1}^{[L]i_L,j_L}) || i_1 \cdots i_L , \, j_1 \cdots j_L \rangle \rangle \,,
  \label{eq:mpo_2}
\end{eqnarray}
where the super-ket $|| i_1 \cdots i_L , \, j_1 \cdots j_L \rangle \rangle = \bigotimes_{a=1}^{L}\ket{i_a}\bra{j_a}$ 
is used in order to deal with the super-operator formalism, i.e. with linear operators 
acting on vector spaces of linear operators.
The bond-link dimension $\chi$ of the MPO~\eqref{eq:mpo_2} can be kept under a given threshold 
by cutting the smallest singular values, and is proportional to the amount of quantum correlations 
between the system sites that can be encoded in $\rho_{\rm \scriptscriptstyle MPO}$.
Starting from $\chi=1$ (separable state) and increasing $\chi$, quantum correlations can be taken 
into account at increasing distance.

The TEBD scheme can be naturally embedded in the Ansatz given in Eq.~\eqref{eq:mpo_2}, 
by performing a Trotter decomposition of the Liouvillian super-operator~\cite{VidalTEBD} 
which describes the master equation~\eqref{eq:Master_Eq} [this can be easily handled 
for Hamiltonian and Lindbladian written as sums of local terms, 
as in Eqs.~\eqref{eq:HamSpin}-\eqref{eq:Lindblad}]. 
In the case of translationally invariant systems, it is even possible to adopt 
an infinite version of the TEBD (the i-TEBD), using the same approach 
that has been successfully applied to pure states~\cite{Vidal_iTEBD}.
Indeed this can be generalized to encompass arbitrary 1D evolution operators 
that can be expressed as a (translationally invariant) tensor network~\cite{Orus2008}.
The TEBD method has been proven to be very effective in many different open 1D quantum systems, 
as for example coupled cavity arrays~\cite{Biella2015,Biondi2015}, Bose-Hubbard chains with bond 
dissipation~\cite{Lauchli2014} and driven/dissipative spin systems~\cite{Joshi2013}.
Alternative approaches based on the variational search of the Liouvillian 
super-operator~\cite{Cui2015, Mascarenhas2015} or on the local 
purification of the density matrix~\cite{Montangero2014} have been recently proposed. 

\begin{figure}[!t]
  \includegraphics[width=0.9\columnwidth]{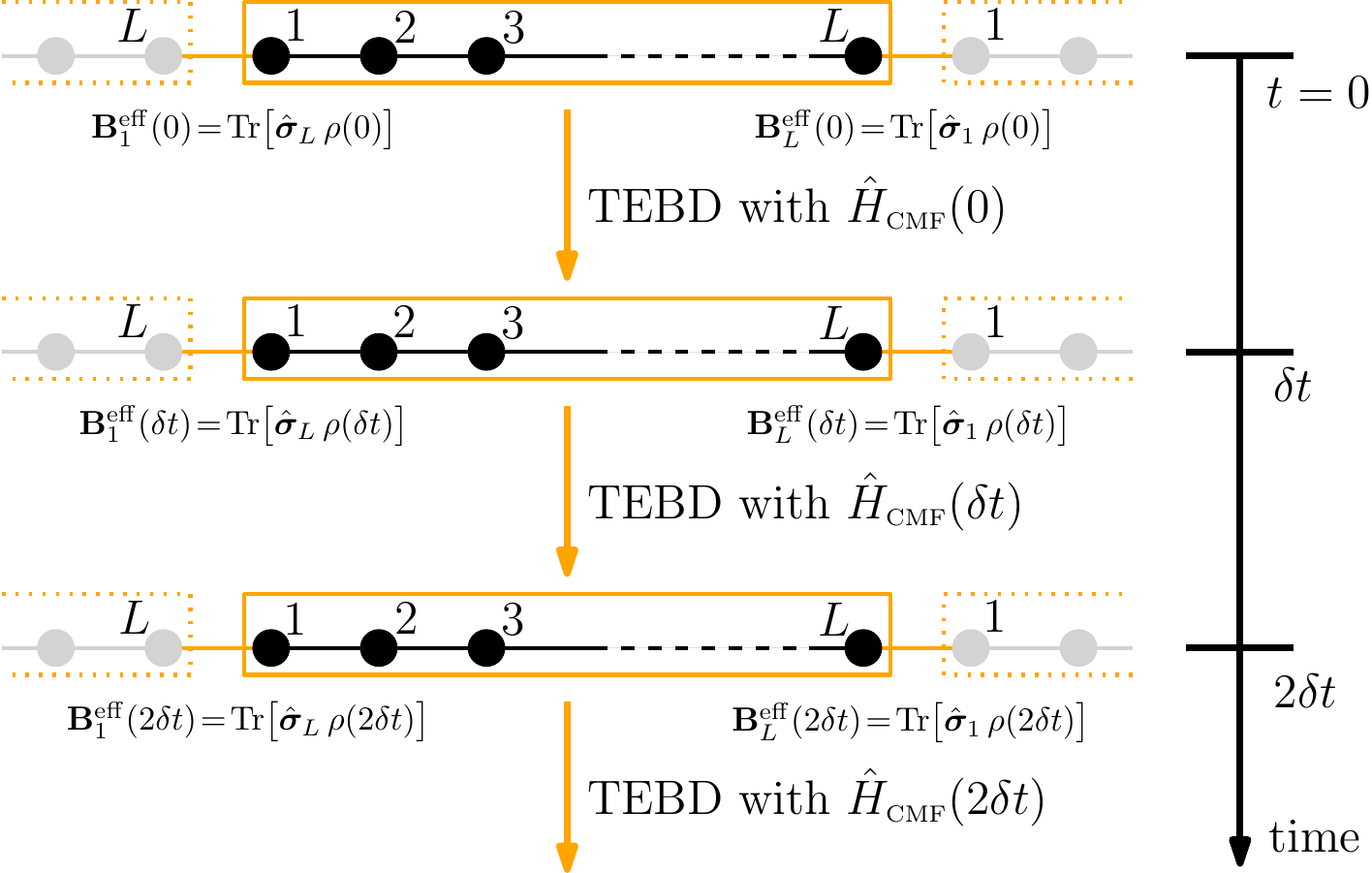}
  \caption{One-dimensional TEBD scheme for 1D systems with open boundaries, combined 
    with the cluster mean-field method. Circles denote the sites of the lattice. 
    The many-body state corresponding to the OBC cluster made up of $L$ black circles 
    inside the orange box is written in an MPO representation, 
    and evolved in time with the TEBD scheme. The cluster is coupled to the rest 
    of the system (gray circles) through the mean field at the edges.
    At regular small time intervals, the mean fields 
    ${\bf B}_1^{\rm eff}(t) = {\rm Tr} \big[ \hat {\boldsymbol \sigma}_L \, \rho(t) \big]$ 
    on the leftmost site and 
    ${\bf B}_L^{\rm eff}(t) = {\rm Tr} \big[ \hat {\boldsymbol \sigma}_1 \, \rho(t) \big]$ 
    on the rightmost site, are self consistently evaluated and used to construct 
    the Hamiltonian for the next TEBD iteration.}
  \label{fig:MPO-CMF}
\end{figure}

The description of 1D dissipative many-body systems in terms of MPO and the search 
for the steady state by time-evolving the Liouvillian super-operator can be combined 
with the CMF approach in a natural way (see Fig.~\ref{fig:MPO-CMF} for a sketch of the idea). 
We consider a linear cluster of $L$ sites with open boundary conditions (OBC); 
its master-equation dynamics can be simulated by means of the TEBD scheme. 
The only novel ingredient is provided by the mean fields which have to be applied 
only at the two edge sites of the chain (the leftmost and the rightmost site). 
These can be easily evaluated in a self consistent way in time, 
by computing the average expectation values: 
\begin{equation}
  {\bf B}^{\rm eff}_1(t) = {\rm Tr} \big[ \hat {\boldsymbol \sigma}_L \, \rho(t) \big] \,, \qquad
  {\bf B}^{\rm eff}_L(t) = {\rm Tr} \big[ \hat {\boldsymbol \sigma}_1 \, \rho(t) \big] \,,
\end{equation}
respectively on site $1$ and site $L$ 
of the chain, at regular time intervals, as outlined above for the other methods. 
Such fields are inserted in the effective Hamiltonian~\eqref{eq:CMF}, which is used to build up 
the Liouvillian operator for the time evolution up to the next iterative step.

As mentioned at the beginning of this subsection, the extension of all these ideas 
to two dimensional systems would be very intriguing. 
For example, one could think to combine the cluster mean-field approach with MPOs
using a mapping of the lattice to a one-dimensional structure with long-range interactions,
through an appropriate \emph{wiring-up} strategy. This has been already successfully
employed in the context of equilibrium systems, where impressive results on wide strips 
have been obtained (see e.g. Ref.~\cite{White2011}).
In higher dimensions, these methods suffer of problems related to the computational cost 
of the tensor network contraction~\cite{Orus2014}, that is common to all planar structures. 
The presence of dissipation could help in reducing the amount of correlations in the steady state, 
so that it might be possible that relatively good accuracies will be reached even with small bond links.

\section{Results}
\label{results}
 
Let us now put into practice the methods outlined above, and study the PM-FM dissipative 
phase transition of the interacting spin model described by Eqs.~\eqref{eq:Master_Eq}, 
\eqref{eq:HamSpin}-\eqref{eq:Lindblad}. As detailed in Sec.~\ref{sec:Model}, this is associated to a 
${\mathbb Z}_2$-symmetry breaking mechanism, whose location in the phase diagram 
we would like to accurately unveil.

The full phase diagram at the single-site MF level has been already obtained in Ref.~\cite{Lee2013}. 
By writing the mean-field equations of motion for the magnetisation along the different axes, 
it is possible to analytically evaluate the critical point separating the PM from the FM phase. 
For fixed values of $J_x, J_z$, it is located at
\begin{equation}
  J_y^c = J_z - \frac{1}{16 \, z^2 \, (J_x-J_z)} \,,
  \label{eq:MF_res}
\end{equation}
where $z$ is the coordination number of the lattice, i.e. the number of nearest neighbors 
of each lattice site. As in any single-site mean field, the only effect of the system dimensionality 
enters through the integer $z$. From the theory of critical phenomena, we know that 
the role of dimensionality is crucial, particularly in low dimensions. 
Below we show that, under a more careful treatment of the short-range correlations,
cluster mean-field produces important qualitative and quantitative changes to the phase diagram. 
In the different subsections, we will address the cases of increasing dimensionality. 
In one-dimensional systems, where we do not expect any phase transition, the cluster mean-field 
together with quantum trajectories and MPO allows to recover this result.

\subsection{One dimension}
\label{sec:1D}

The 1D case represents the most suitable situation to benchmark our methods.
Here, due to the reduced dimensionality (the system has a coordination number $z=2$),
the MF predictions are known to fail and no symmetry-breaking mechanism should occur
(as already stated in Ref.~\cite{Lee2013}).
Using a combination of strategies as described in Sec.~\ref{sec:methods}, we numerically verify 
the absence of symmetry breaking, thus gaining confidence on how accurate our methods 
can be for driven-dissipative systems.

We are able to perform a direct integration of the master equation~\eqref{eq:Master_Eq} for systems 
with up to $L=9$ spins, by employing a standard fourth-order Runge-Kutta (RK) method,
without applying consistent MF terms at the boundaries. 
For larger systems, with $10 \leq L \leq 16$, we use the quantum trajectory approach 
(the time evolution of each trajectory is computed by means of a fourth-order RK method)
obtaining reliable results already with a number of trajectories not exceeding $N = 500$, 
for all the values of the parameters we have probed. For even larger clusters ($L \lesssim 40$) 
we resort to an MPO approach combined with the cluster mean field.

In order to check for the (possible) existence of an ordered FM phase, 
we calculate the steady-state ferromagnetic spin-structure factor 
$S^{xx}_{\scriptscriptstyle \rm SS} (k=0)$, where
\begin{equation}
  S^{xx}_{\scriptscriptstyle \rm SS} (k) = 
  \frac{1}{L^2} \sum_{j,l=1}^L e^{-ik(j-l)} \langle \sigma^x_j \sigma^x_l \rangle_{\scriptscriptstyle \rm SS} \,.
  \label{eq:xx_Correl}
\end{equation}
A non-zero value of $S^{xx}_{\scriptscriptstyle \rm SS}(0)$ indicates the stabilization of a FM ordering 
in the thermodynamic limit. We do not look directly at the order parameter 
$\langle \sigma^x_j \rangle_{\scriptscriptstyle \rm SS}$, since we are studying finite-size systems 
and the ${\mathbb Z}_2$ symmetry may not spontaneously break~\cite{note02}.

\begin{figure}[!t]
  \includegraphics[width=0.98\columnwidth]{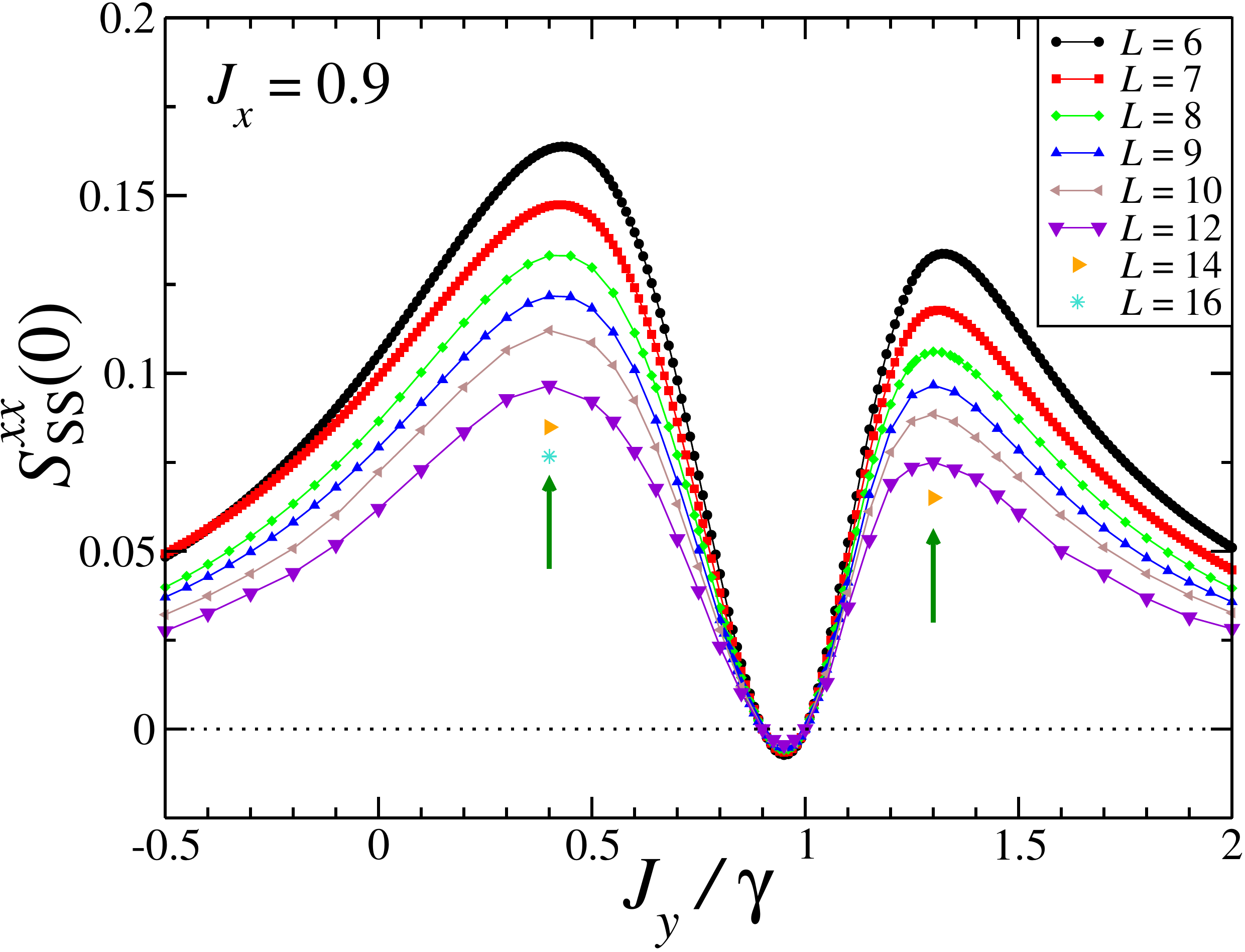}
  \caption{Ferromagnetic spin-structure factor along the $x$ direction, 
    in a 1D setup, as a function of $J_y$. 
    The various curves and symbols stand for different
    system sizes from $L=6$ to $L=16$, as indicated in the plot.
    The two arrows point at the positions of the two peaks ($J_y=0.4$ and $J_y=1.3$),
    for which we provide a finite-size scaling (Fig.~\ref{fig:1D_Scaling}), 
    and an analysis of the two-point correlation functions (Fig.~\ref{fig:1D_Correl-R}).
    Here we have set $J_x = 0.9$ and $J_z = 1$, and work in units of $\gamma$.
    Note that for $J_y=0.9$ and $J_y=1$ the spin structure factor is rigorously zero.}
  \label{fig:1D_Correl}
\end{figure}

In Fig.~\ref{fig:1D_Correl} we show the behavior of $S^{xx}_{\scriptscriptstyle \rm SS}(0)$ 
for small systems ($L\leq 16$) with open boundary conditions,
for fixed values of $J_x=0.9, \, J_z=1$ and varying $J_y$
(analogous results are obtained by taking different values of $-1 \leq J_x \leq 1$). 
Data have been obtained with RK and with QT approaches.
According to Eq.~\eqref{eq:MF_res}, the MF approach predicts a critical point 
at $J_y^c = 37/32 \approx 1.156$ separating a PM (for $J_y < J_y^c$) 
from a FM region (for $J_y > J_y^c$).
In striking contrast with this, our numerics displays a decrease of the $xx$ correlations 
with the system size.
We also observe a non-monotonic behavior with $J_y$, and the fact that 
$S^{xx}_{\scriptscriptstyle \rm SS}(0)$ vanishes for $J_y=0.9$ and $J_y=1$.
This can be explained as follows.
For $J_x = J_y$, the Hamiltonian~\eqref{eq:HamSpin} conserves the magnetization
along $z$. Since the dissipative spin-flip processes occur along the same direction, 
it is clear that they cannot be counteracted by the unitary dynamics and so the steady state 
is a pure product state, having all the spins aligned and pointing down along the $z$ direction,
making the $xx$ and $yy$ correlations vanishing at any distance.
On the contrary for $J_y = J_z$, the total magnetization along the $x$-axis is conserved by the Hamiltonian.
In this case, due to the different privileged axis with respect to the dissipation process, 
the steady state is not a product state. The correlators are generally different from
zero, however the spin-structure factor of Eq.~\eqref{eq:xx_Correl} at $k=0$ sums to zero.
It is worth noticing that, on the contrary, $S^{yy}_{\rm \scriptscriptstyle SS}(k=0)$ 
is not affected by the Hamiltonian symmetry and is different from zero (not shown).

\begin{figure}[!t]
  \includegraphics[width=0.99\columnwidth]{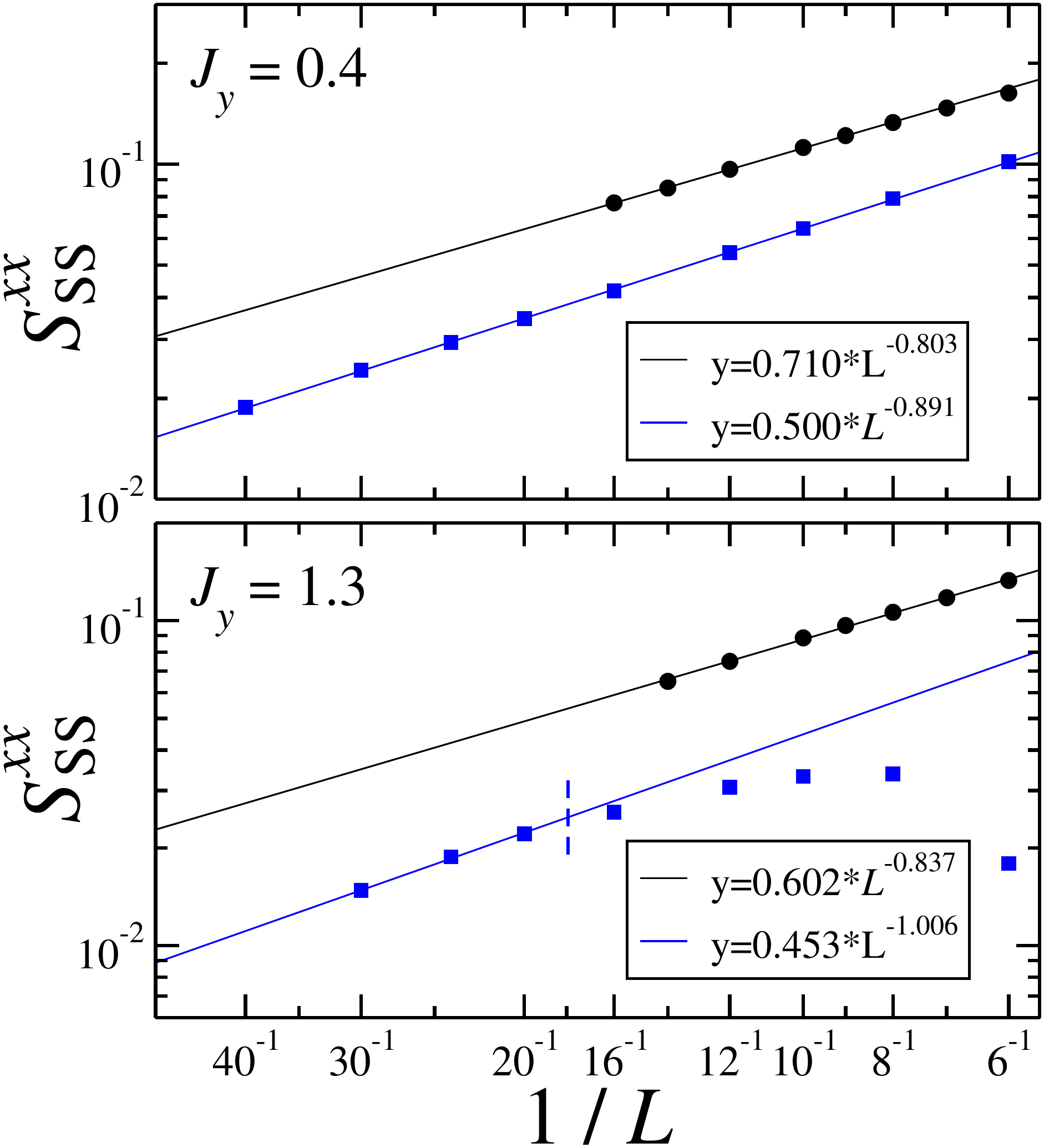}
  \caption{Scaling of $S^{xx}_{\scriptscriptstyle \rm SS}(0)$ as a function 
    of the inverse system size $L$, for two values of $J_y$ in proximity of the peaks
    (see the arrows in Fig.~\ref{fig:1D_Correl}). 
    The symbols denote the numerical data, while the continuous lines
    are power-law fits performed for the data points to the left of the vertical dashed line.
    The black sets correspond to those of Fig.~\ref{fig:1D_Correl},
    obtained simulating a small system with RK and QT approaches.
    The blue sets have been obtained with MPO simulations, where the CMF has been 
    applied to the two edges. The other parameters are set as in Fig.~\ref{fig:1D_Correl}.}
  \label{fig:1D_Scaling}
\end{figure}

Coming back to the data in Fig.~\ref{fig:1D_Correl} on the spin-structure factor
$S^{xx}_{\rm \scriptscriptstyle SS}(k=0)$, we can pinpoint the emergence of two peaks 
at $J_y \approx 0.4$ and $1.3$.
Before commenting on the behavior of the spin-structure factor in proximity of such peaks, 
let us analyze more in detail their dependence with $L$ by performing a finite-size 
scaling of our data. This is provided in Fig.~\ref{fig:1D_Scaling}.
Black data sets correspond to those in Fig.~\ref{fig:1D_Correl}.
We observe a systematically drop of the correlations with $L$
for both peaks, that can be nicely fitted with a power law
\begin{equation}
  S^{xx}_{\scriptscriptstyle \rm SS}(0) \sim \kappa \, L^{-\alpha} \,,
  \label{eq:Sxx_powerlaw}
\end{equation}
where the exponent $\alpha$ depends on the value of $J_y$ as indicated in the various panels. 

We were able to reach longer sizes by employing a MPO approach for considerably larger 
chain lengths ($L \leq 40$). We applied a cluster mean field at the edges of the chain, 
in order to better mimic the thermodynamic limit. The results obtained with this method are 
displayed in Fig.~\ref{fig:1D_Scaling} by the blue sets of data, and qualitative agree 
with the previous results without mean field (black data). 
In particular, an analogous power-law behavior~\eqref{eq:Sxx_powerlaw} emerges. 
Notice that, in correspondence to the peak that is remnant of the ferromagnetic phase
($J_y=1.3$), a non-monotonic behavior in the combined MPO-CMF approach emerges.
This has to be ascribed to the mean-field corrections that become very effective for 
very short clusters.

\begin{figure}[!t]
  \includegraphics[width=0.98\columnwidth]{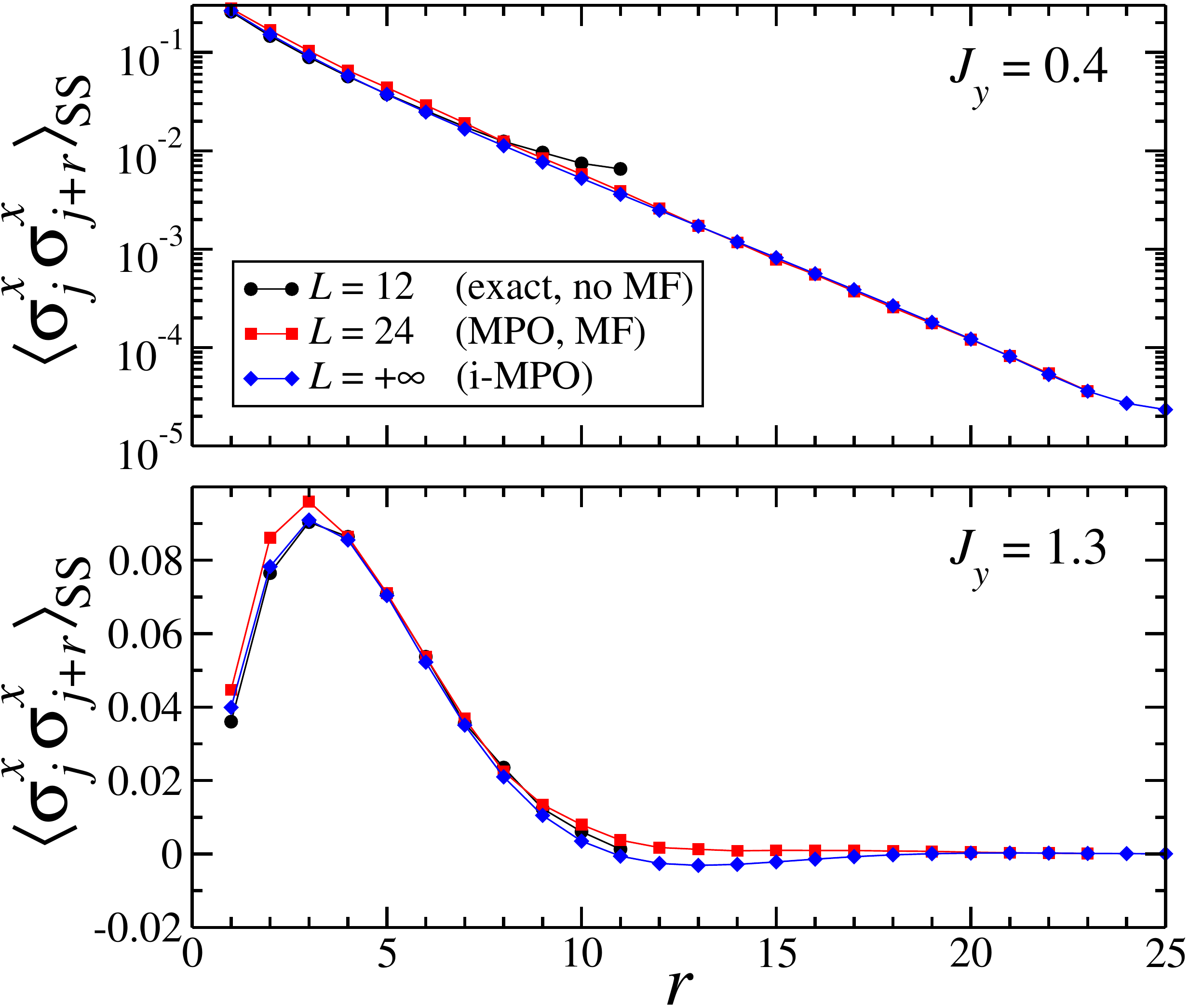}
  \caption{Spatial decay of the correlation functions 
    $\langle \sigma^x_j \sigma^x_{j+r} \rangle_{\scriptscriptstyle \rm SS}$ 
    with the distance $r$. Correlators have been chosen in a symmetric way 
    with respect to the center of the chain.
    In the upper panel $J_y = 0.4$ (left peak in Fig.~\ref{fig:1D_Correl}),
    while in the lower panel $J_y = 1.3$ (right peak in Fig.~\ref{fig:1D_Correl}).
    The various data sets correspond to different system sizes:
    results for $L = 12$ have been obtained for systems with PBC
    by means of RK integration or QT approach to the master equation;
    those for $L=24$ are with MPO used to simulate OBC and combined with the CMF;
    the thermodynamic limit $L \to \infty$ (diamonds--solid blue lines)
    has been addressed with a translationally invariant i-MPO method.
    The other parameter values are set as in Fig.~\ref{fig:1D_Correl}.}
  \label{fig:1D_Correl-R}
\end{figure}

Further evidence of the remnants of the ${\mathbb Z}_2$-symmetry breaking predicted 
at the mean-field level is provided by analyzing the two-point correlation functions 
$\langle \sigma^x_j \sigma^x_{j+r} \rangle_{\scriptscriptstyle \rm SS}$ as a function of the distance $r$.
Figure~\ref{fig:1D_Correl-R} shows results for parameters corresponding to the two distinct phases 
predicted by the mean-field theory. 
In particular we observe that, in the cases where the symmetry is not broken in MF, 
correlations of the order parameter exhibit a clear exponential decay with the distance, 
as one can recognize in the upper panel ($J_y = 0.4$). 
This is evident already at very small sizes $L \sim 12$. 
A more intriguing situation occurs in the case where the MF would predict 
a symmetry-broken phase (see the lower panel for $J_y = 1.3$). 
In such case an instability of the PM phase at short lengths emerges, 
in the sense that a bump in the correlators clearly emerges at $r \lesssim 10$
and the exponential suppression of correlations is not immediately visible.
Longer sizes are needed to observe the absence of quasi-long-range correlations.

To corroborate our analysis, we also performed simulations by directly addressing 
the thermodynamic limit.
We employed a TEBD numerical approach based on a translationally invariant Ansatz 
for the MPO~\cite{Orus2008}. Here the mean field need not to be used.
The results are in perfect agreement with those obtained with the cluster mean field,
thus validating our approach.
In all the cases that we considered, we clearly see an emergence of exponential decay at large distances,
thus signalling the absence of ferromagnetic order in any parameter range.
Remarkably, the data obtained with MPO simulations (both in the finite and the infinite case), 
converge with a relatively small bond-link dimension ($\chi\le120$).

\subsection{Two dimensions}
\label{sec:2D}

We now proceed with the discussion of the model in a two-dimensional square lattice ($z=4$). 
Here there is no chance to solve Eq.~\eqref{eq:Master_Eq} exactly for any 
thermodynamically relevant system size, therefore we resort to approximate techniques 
combined with a CMF approach.
In this framework we are able to highlight a number of significant modifications 
to the steady-state phase diagram predicted by the single-site MF. 
Clearly such differences must arise from taking into account the effect 
of short-range correlations inside the cluster.
The shape of the considered clusters always respects the square-lattice geometry 
(i.e. they have a number of sites $L = \ell \times \ell$). 
With the numerical capabilities at our disposal, we are able to deal 
with clusters up to size $\ell = 4$.
The $\ell \leq 3$ data have been computed by numerically integrating 
the time evolution of the cluster master equation with a standard RK method. 
In order to address the case $\ell = 4$, we employed the quantum trajectories approach 
explained in Sec.~\ref{subsec:qt}.

\begin{figure}[!t]
  \includegraphics[width=0.99\columnwidth]{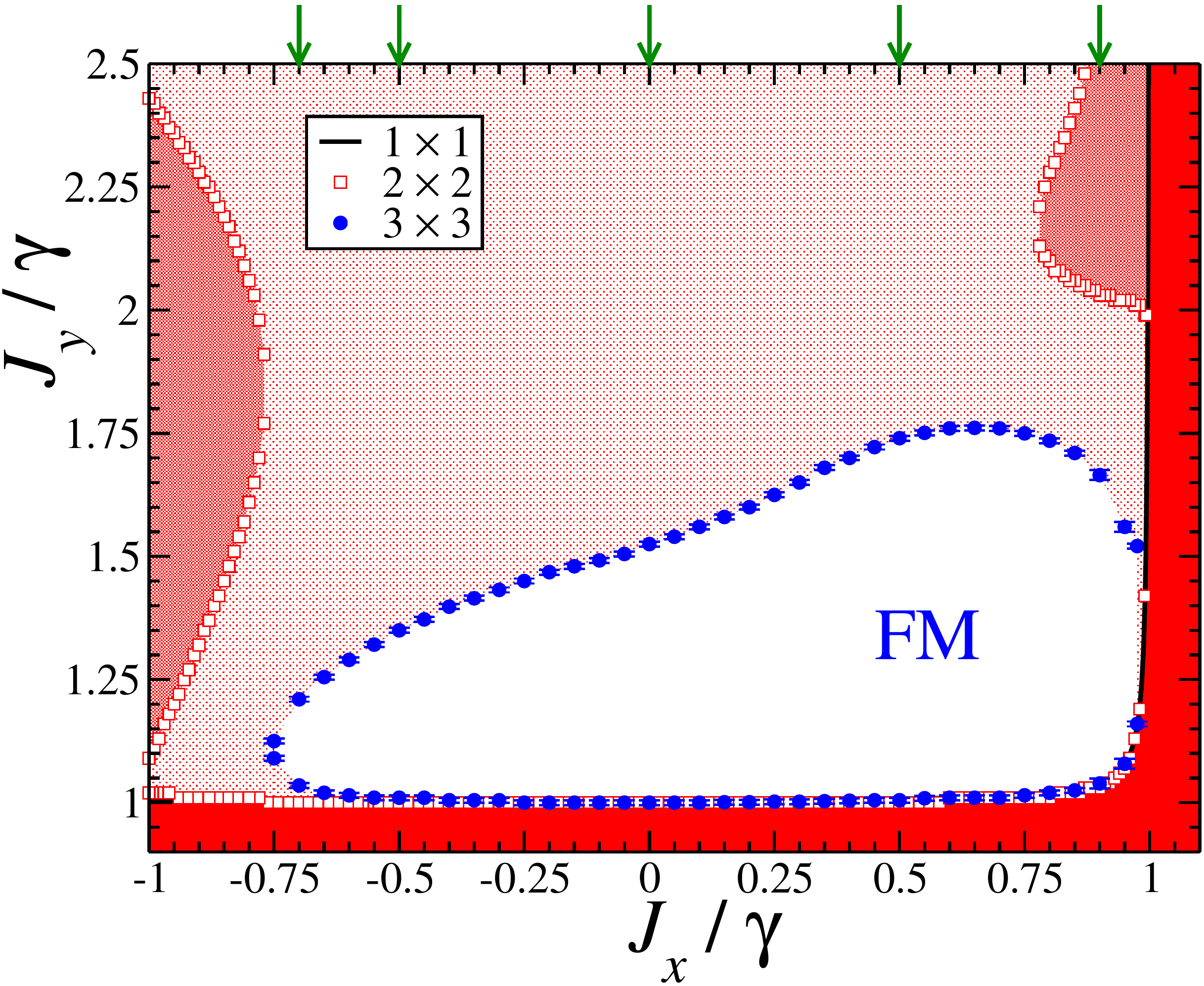}
  \caption{Two-dimensional cluster mean-field phase diagram 
    in the $J_x$-$J_y$ plane (with fixed $J_z = 1$).
    The single-site MF ($1 \times 1$) predicts a ferromagnetic steady state
    in the top left region with respect to the black curve.
    The extent of this region appears to be very fragile to a more accurate
    cluster mean-field treatment.
    At the $2 \times 2$ level (red squares), the boundaries of the two phases 
    are slightly deformed, while with a $3 \times 3$ analysis 
    the FM phase shrinks down to a region of finite size (blue circles).
    The darkest color filling indicates the region that is PM in all simulations, 
    while the lightest indicates that which is FM in all cases. 
    The FM region shrinks with increasing cluster
    size, as indicated by the varying shades of color, but, as discussed
    in the text, appears to converge with increasing cluster size.
    The five arrows denote the cuts along different values of $J_x$ 
    which will be analyzed in detail in Figs.~\ref{fig:2D_Cuts} and~\ref{fig:2D_Cut09}.}
  \label{fig:2D_PhaseDiag}
\end{figure}

Our main result is reported in Fig.~\ref{fig:2D_PhaseDiag}, which displays the phase diagram 
in a region of the parameter space where the MF analysis would predict the occurrence of a 
${\mathbb Z}_2$-symmetry breaking mechanism. 
It is immediately visible that, under a CMF treatment of the system, the extent
of the FM phase is drastically reduced.
Specifically we shall contrast the single-site MF predictions (black line) with the results 
obtained using a $3 \times 3$ cluster size (blue circles).
On the one side, the single-site MF analysis predicts a symmetry-broken phase 
in a large and extended portion of the phase space 
[fixing $J_z=1$, for $-1 \lesssim J_x \lesssim 1$ the ferromagnet extends 
for any $J_y \gtrsim 1$ according to Eq.~\eqref{eq:MF_res}, and disappears only
in the asymptotic limit $J_y \to \infty$].
On the other side, the latter analysis indicates a tendency to confine the FM phase 
into a finite-size region in the parameter phase, which is surrounded by the PM phase, 
thus modifying the topology of the diagram. 

Our CMF numerics shows that the disappearance of the ordered phase at large $J_y$
is accompanied by the progressive shrinkage of the Bloch vector 
for the single-site density matrix, with increasing the coupling strength.
This effect can be already seen from the Bloch equations of the single-site MF~\cite{Lee2013}, 
which predict a saturation of the spins in the limit 
of infinite coupling---see Eqs.~\eqref{eq:mx_1site}, \eqref{eq:mz_1site} 
and analogous for $[M^y_{\rm \scriptscriptstyle SS}]_{\rm \scriptscriptstyle MF}$.
It is however important to remark that, even though the left and upper boundaries of the FM phase 
shrink with the cluster size while the right and bottom ones are almost unaffected,
our results support the existence of a finite region for the symmetry-broken phase 
even in the thermodynamic limit $\ell \to \infty$, as we will detail below.
Since the calculations with large clusters are very demanding, 
we considered few (representative) couplings.
Our analysis performed with clusters of size up to $4 \times 4$ indicates that 
the ferromagnet will survive in the limit $\ell \to \infty$, for fixed $J_x = 0.9$
and for $1.04 \lesssim J_y \lesssim 1.4$ (see Fig.~\ref{fig:2D_Cut09}). We do expect that for 
other values of $J_x$ the behavior will be similar.

Before commenting on the scaling with the cluster size, let us point out the fact that 
the CMF data for $\ell = 2$ (red squares in Fig.~\ref{fig:2D_PhaseDiag}) evidence
an intermediate situation. Indeed taking into account only nearest-neighbor interactions,
the extent of the FM phase is slightly reduced as compared to the single-site MF, 
yet it is not sufficient to confine the symmetry-broken phase into a finite-size region 
surrounded by the PM phase.
Nonetheless, after a more careful analysis of the magnitude of the order parameter,
we are able to detect a clear tendency toward a topological modification of the diagram.
Specifically we fixed several values of the coupling $J_x$, while varying $J_y$,
and investigated the FM-PM phase transition by looking at the steady-state 
on-site magnetisation along the $x$-axis:
\begin{equation}
  M^x_{\rm \scriptscriptstyle SS} = \frac{1}{\ell^2} \sum_{j=1}^{\ell^2} \langle \sigma^x_j \rangle_{\rm \scriptscriptstyle SS} \,,
\end{equation}
so to explore the phase diagram of Fig.~\ref{fig:2D_PhaseDiag} along certain vertical cuts.
Notice that we do not need to calculate the correlators $S^{xx}_{\rm \scriptscriptstyle SS}(0)$ 
of Eq.~\eqref{eq:xx_Correl} as we did in the 1D geometry, since the self-adaptive 
mean-field method automatically breaks the symmetry in the FM phase. 

\begin{figure}[!t]
  \includegraphics[width=0.95\columnwidth]{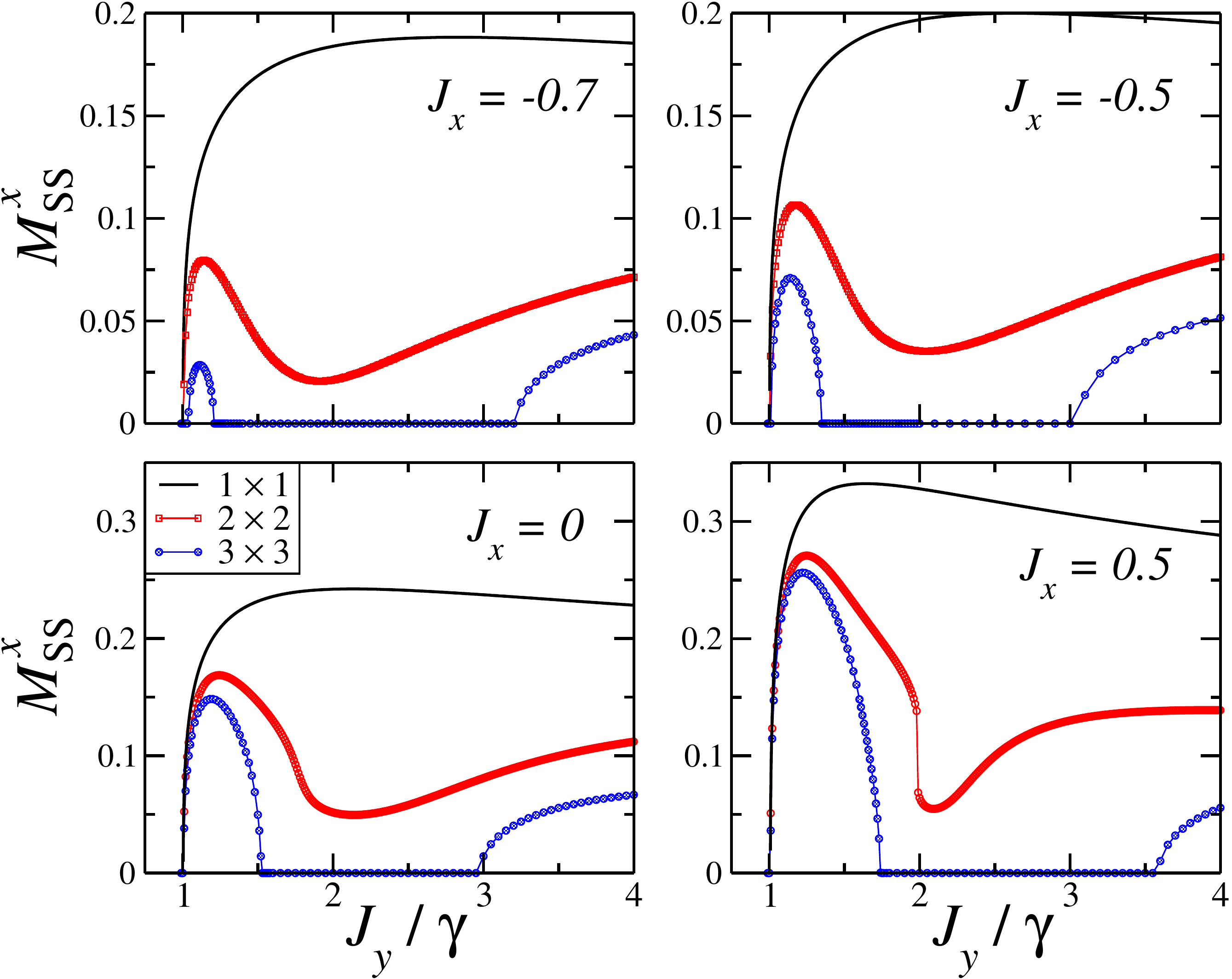}
  \caption{Cluster mean field analysis of the ferromagnetic order parameter 
    in two dimensions, for four different vertical cuts of Fig.~\ref{fig:2D_PhaseDiag}, 
    at constant $J_x$ (the corresponding values of $J_x$ are indicated in the various panels).
    The various data sets denote different sizes of the clusters, up to $\ell = 3$.}
  \label{fig:2D_Cuts}
\end{figure}

The different panels of Fig.~\ref{fig:2D_Cuts} refer to four values of $J_x$, as indicated
by the first four green arrows on the left in Fig.~\ref{fig:2D_PhaseDiag}, 
and display $M^x_{\rm \scriptscriptstyle SS}$ as a function of $J_y$, for different cluster sizes $\ell$.
The $1 \times 1$ MF data (black lines) can be found by working out the steady-state limit of the MF
Bloch equations for the magnetization~\cite{Lee2013}, giving the following result:
\begin{equation}
  [M^x_{\rm \scriptscriptstyle SS}]_{\rm \scriptscriptstyle MF} = 
  \pm \sqrt{2 \, [M^z_{\rm \scriptscriptstyle SS}]_{\rm \scriptscriptstyle MF} \Big( [M^z_{\rm \scriptscriptstyle SS}]_{\rm \scriptscriptstyle MF} +1 \Big) \frac{J_y-J_z}{J_x-J_y} } \,,
  \label{eq:mx_1site}
\end{equation}
with
\begin{equation}
  [M^z_{\rm \scriptscriptstyle SS}]_{\rm \scriptscriptstyle MF} =
  - \frac{1}{4z} \sqrt{\frac{1}{(J_z-J_x)(J_y-J_z)}} \,.
  \label{eq:mz_1site}
\end{equation}
These curves exhibit a finite magnetization for all $J_y \gtrsim 1$, with a maximum at a given 
value of $J_y$ (dependent of $J_x$) and they eventually go to zero in the limit $J_y \to +\infty$.
This vanishing order at strong coupling is similar to the
absence of ordering on resonance in the Dicke model~\cite{baumann}, and
the suppression of ordering in the degenerate limit of the Rabi model~\cite{Schiro2015}.
The non-monotonicity of $M^x_{\rm \scriptscriptstyle SS}$ as a function of $J_y$
emerges also in the CMF analysis: the $2 \times 2$ data signal a strong suppression 
of the order parameter for $J_y \sim 2$, which however remains finite.
Going further with a $3 \times 3$ cluster, we see the sharp disappearance of the FM 
in an intermediate extended region where $M^x_{\rm \scriptscriptstyle SS} = 0$
(for $1.5 \lesssim J_y \lesssim 3$, depending on the value of $J_x$,
the system is not ferromagnetically ordered along $x$ or $y$).
The revival of the FM phase at large values of $J_y$ ($J_y \gtrsim 3$)
is outside the parameter range of Fig.~\ref{fig:2D_PhaseDiag}.
We will analyze this feature later in Sec.~\ref{sec:stability}.

\begin{figure}[!t]
  \includegraphics[width=0.95\columnwidth]{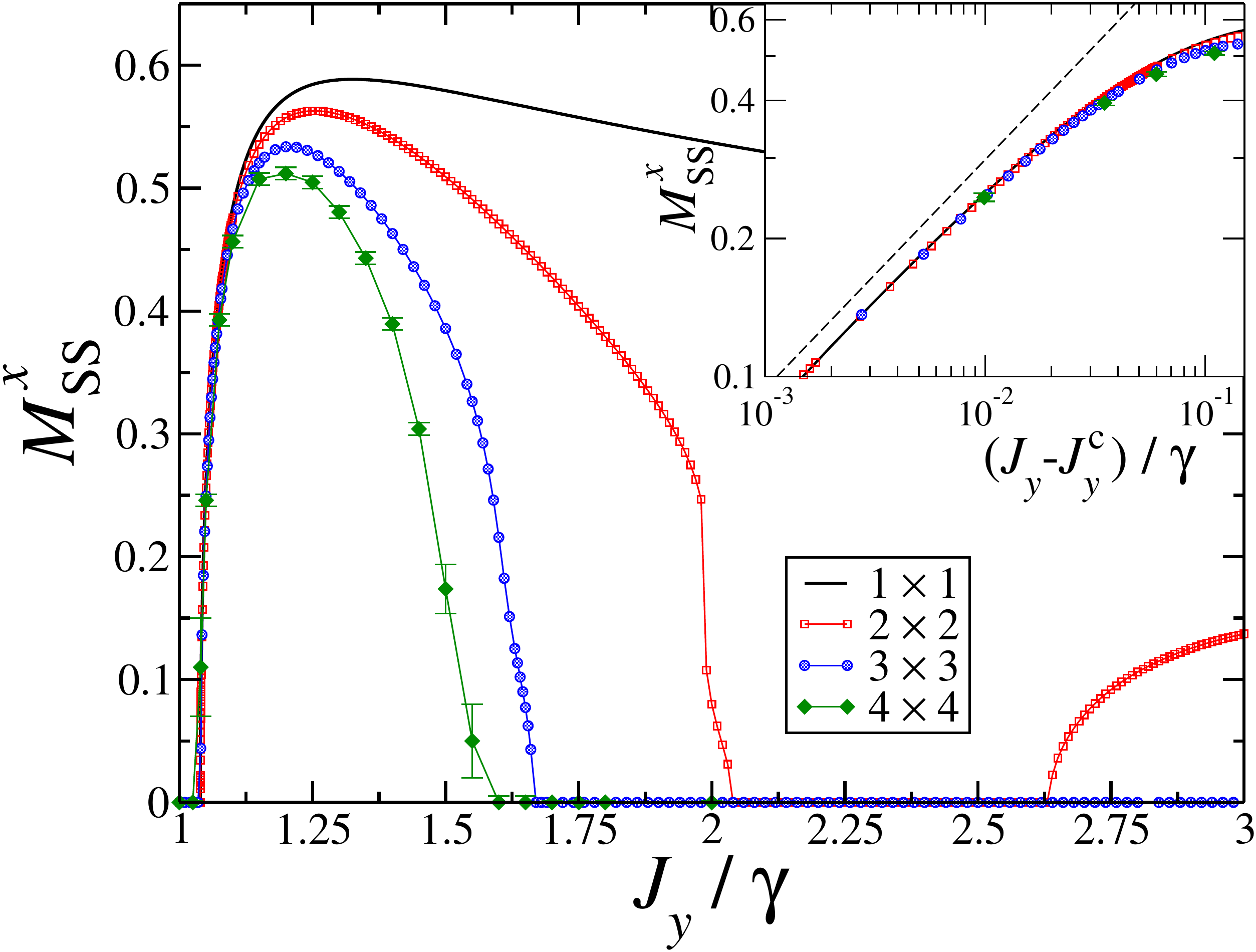}
  \caption{Same analysis as in Fig.~\ref{fig:2D_Cuts}, but for $J_x$ = 0.9. 
    In the inset we show the rescaled data close to the left phase transition;
    the straight dashed line denotes a square-root behavior as in Eq.~\eqref{eq:Mx_mf},
    and is plotted as a guide to the eye.}
  \label{fig:2D_Cut09}
\end{figure}

Let us now have a closer look at the vertical cut of Fig.~\ref{fig:2D_PhaseDiag} for $J_x=0.9$;
the magnetization is shown in Fig.~\ref{fig:2D_Cut09} for clusters up to $\ell = 4$. 
Both for the $3 \times 3$ and the $4 \times 4$ CMF analysis, we do not see any reappearance 
of the FM ordering at large $J_y$ (we numerically checked this statement up to $J_y = 10$).
The symmetry-broken phase is confined to a finite-size region which shrinks with increasing $\ell$.
While the left boundary is basically unaffected by the role of correlations
($J_y^{c \, ({\rm left})} \approx 1.04 \pm 0.01$), the right boundary is strongly sensitive to $\ell$.
Our simulations indicate a transition point $J_y^{c ({\rm right})} \approx 2.04 \pm 0.005, \,
1.67 \pm 0.01, \, 1.57 \pm 0.03$, for clusters respectively with $\ell = 2, 3, 4$. 
A scaling with $\ell$ of these data for the right boundary 
indicates a behavior that is compatible with $J_y^{c \, ({\rm right})} \approx 1.40 + 2.54 \, \ell^{-2}$, 
thus which supports the existence of the FM phase in the limit of large cluster size $\ell \to \infty$,
for $1.04 \lesssim J_y \lesssim 1.40$.
In the data for $\ell = 2$, a discontinuity of $M^x_{\rm \scriptscriptstyle SS}$ seem to appear
immediately before the right transition point (at $J_y \approx 2$), which requires
a further analysis (a similar behavior is observed in the lower right panel of Fig.~\ref{fig:2D_Cuts},
for $J_y = 0.5$). We will return to this point in Sec.~\ref{sec:stability}.

We also checked that, close to the transition, our numerics predicts a growth of the order parameter 
that is well approximated by
\begin{equation}
  M^x_{\rm \scriptscriptstyle SS} \sim m \sqrt{J_y-J_y^c} \,,
  \label{eq:Mx_mf}
\end{equation}
as displayed in the inset of Fig.~\ref{fig:2D_Cut09}, around the left critical point $J_y^{c\, ({\rm left)}}$. 
We repeated a similar analysis for other vertical (fixed $J_y$) 
and horizontal (fixed $J_x$) cuts, and obtained qualitatively analogous results.
This evidences the fact that the CMF remains a mean-field analysis, and leads to 
the same critical exponents as those of its single-site version. 
In order to get the correct exponents, one would need a more careful finite-size
analysis~\cite{Suzuki1986} which requires slightly larger values of $\ell$
and is unfortunately out of reach for the present computational capabilities.

\begin{figure}[!t]
  \includegraphics[width=0.98\columnwidth]{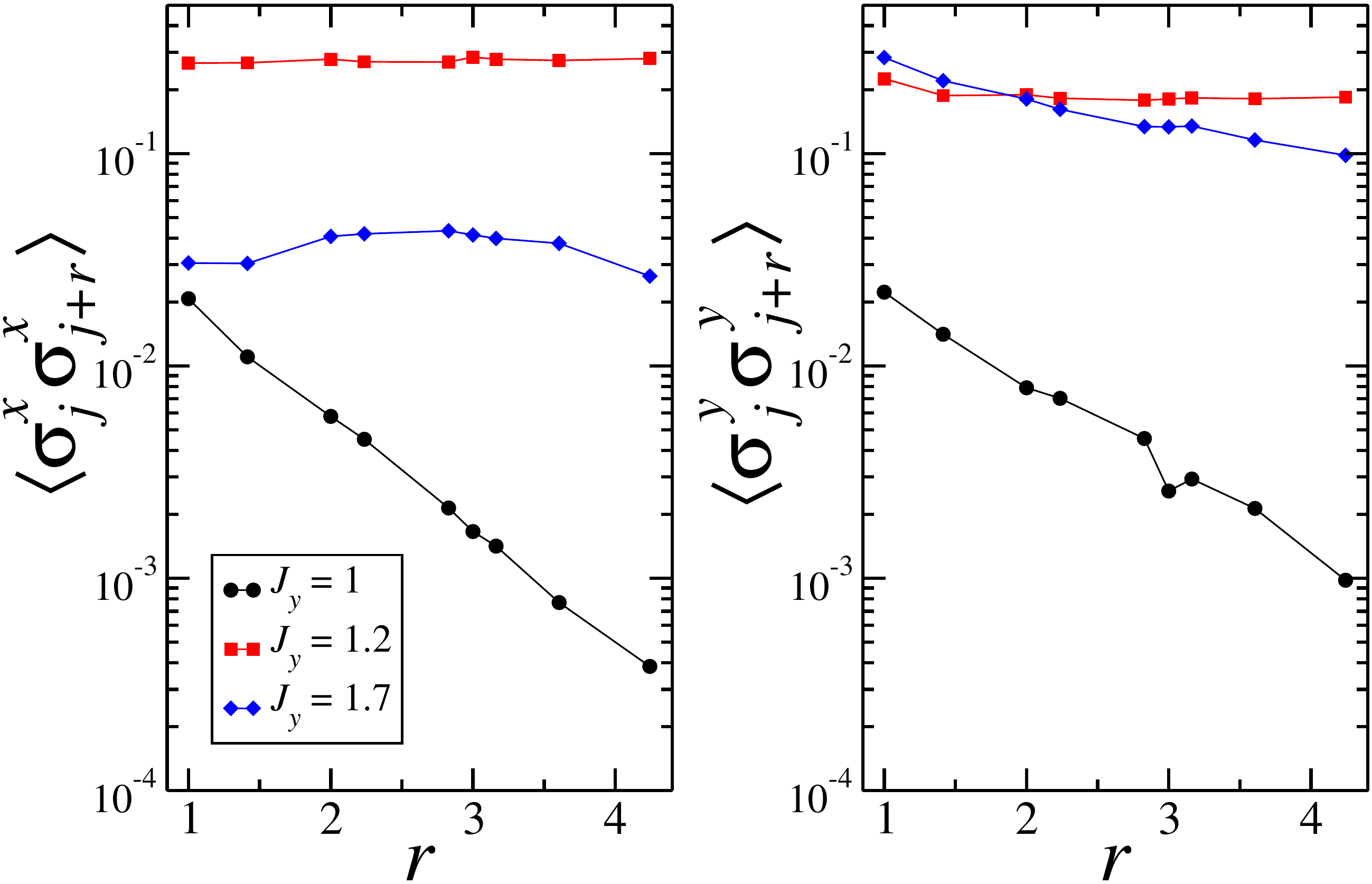}
  \caption{The two-point $xx$ (left) and $yy$ (right) 
    correlations as a function of the distance $r$ in a two-dimensional 
    square-lattice geometry. The calculations have been performed 
    on a square lattice of size $\ell=4$, while the point $j$ 
    has been chosen to be at one of the corners of the squarer cluster. 
    The three sets of data refer to different values of $J_y$ according
    to the legend: two inside the PM phase ($J_y= 1$ and $J_y=1.7$) and
    one inside the FM phase ($J_y=1.2$).
    We fixed $J_x = 0.9$, $J_z = 1$.}
  \label{fig:Corr2D}
\end{figure}

The stability of the symmetry-broken phase for $J_y^{c \, ({\rm left})} < J_y < J_y^{c \, ({\rm right})}$
up to $4 \times 4$ clusters is corroborated by the behavior of the correlation functions 
$\langle \sigma^x_j \sigma^x_{j+r} \rangle_{\rm \scriptscriptstyle SS}$ 
and $\langle \sigma^y_j \sigma^y_{j+r} \rangle_{\rm \scriptscriptstyle SS}$ with the distance $r$, 
as reported in Fig.~\ref{fig:Corr2D} for three different values of $J_y$. 
As discussed in Section~\ref{sec:1D} for the 1D case, in the parameter region where we predict a PM, 
the correlators decay exponentially with $r$ (black data set at $J_y = 1$). 
On the opposite side, the point at $J_y=1.2$ (red data) displays a marked distance-independence 
of correlations with the distance, thus signalling the presence of a FM phase (notice that 
this point lies well inside the closed region in Fig.~\ref{fig:2D_PhaseDiag}).
The case $J_y=1.7$ (blue data) shows a subtler behavior and corresponds to a point for which 
the single-site and the $2 \times 2$ mean-field analysis would predict a symmetry-broken phase, 
contrary to our $\ell \geq 3$ CMF calculations which display no evidence of this type. 
The reminiscence of a kind of quasi-ordering 
at short distances is indeed forecast by a slow decay of correlations. 
While we are not able to see a clear exponential decay with $r$, due to our limited 
numerical capabilities, we expect that this would be visible for clusters appreciably 
longer than $\ell = 4$. Nonetheless we stress that $xx$ correlations here are 
one order of magnitude smaller than in the FM point.

A sketch of the phase diagram summarising all our results is provided in Fig.~\ref{fig:diagram}.

\subsection{Two dimensions - stability analysis}
\label{sec:stability}

As anticipated in the previous subsection, the $2 \times 2$ analysis reveals
a discontinuity of the order parameter inside the first FM phase, 
very close to the transition point $J_y^{c \, {\rm (right)}}$ to the disordered phase.
Such a jump, between two symmetry-broken states, is known as a metamagnetic transition.
The jump is visible for certain values of $J_x$, and seems to vanish quickly
with increasing the cluster size (already for $\ell = 3$ it is barely recognizable from our data).
On the one hand, the latter observation suggests that this jump could be an artifact 
of the CMF analysis. On the other hand, a deeper investigation is required
to understand its origins.

\begin{figure}[!b]
  \includegraphics[width=0.9\columnwidth]{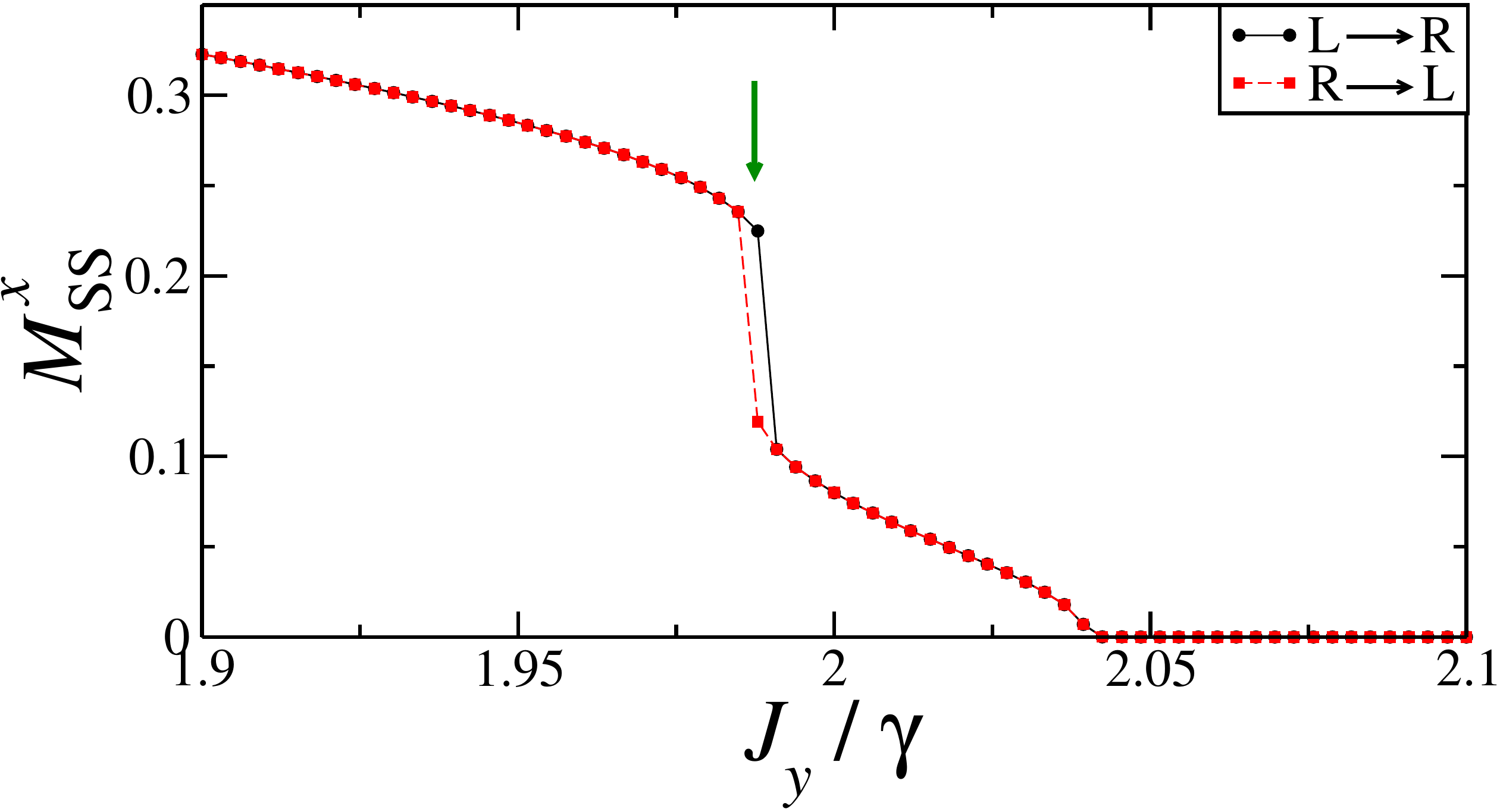}
  \caption{Magnification of the $2 \times 2$ CMF analysis for $J_x=0.9$, close to $J_y=2$
    (see the red data in Fig.~\ref{fig:2D_Cut09}).
    The two sets are calculated sweeping both from upward and downward $J_y$ values, 
    where the initial conditions for each point are based on previous one, with a small offset. 
    This evidences the presence of a first-order jump within the ordered phase
    for $J_y \approx 1.985 $ (green arrow), 
    followed by a second order transition to the normal state, for $J_y^{c\,({\rm right})} \approx 2.04$.}
  \label{fig:2x2_jump}
\end{figure}

To highlight the existence of this feature, in Fig.~\ref{fig:2x2_jump} 
we show a magnification of the relevant parameter region of Fig.~\ref{fig:2D_Cut09}.
We only consider the $2 \times 2$ case, since this is the situation where it is mostly relevant.
We observe the presence of a first-order phase transition within the first ordered phase, 
where the order parameter exhibits a discontinuity. This is corroborated by a bistability effect:
specifically, we calculated the magnetization $M^x_{\rm \scriptscriptstyle SS}$ 
starting from different initial states and we observed a slight difference in proximity
to the jump, as is visible from the figure~\cite{note03}.

At this point we perform a linear stability analysis, 
in order to check whether and how the system becomes unstable in correspondence of the jump.
We start from the CMF factorization Ansatz given in Eq.~\eqref{eq:rho_cmf},
where each cluster density matrix $\rho_{\mathcal C}$ obeys the mean-field 
master equation~\eqref{eq:Master_cluster}.
The stability analysis is performed directly on the factorized density matrix, 
as detailed in Ref.~\cite{LeBoite2013}.
Let us first rewrite the equation of motion for a single cluster, say the $n$-th one,
in the superoperator formalism as:
\begin{equation}
  \partial_t \sket{\rho_n} = \mathcal{M}_0 \sket{\rho_n}
  + \sum_j \Big( \mathcal{E}_j\cdot \sket{\rho_{n+\bf{e_j}}} \Big) \, \mathcal{M}_j \sket{\rho_n} \,,
\end{equation}
where we omitted the index ${}_{\mathcal C}$.
Here $\sket{\rho_n}$ denotes a super ket, i.e.. a vectorised form
of the density matrix, and $\mathcal{M}_i$ denote superoperators.
In this equation $\mathcal{M}_0$ represents all the on-cluster terms, 
while $\mathcal{M}_j$ is the on-cluster part of an off-cluster term, 
and $\mathcal{E}_j$ the corresponding off-cluster expectation.
For example, in the term $J_x \langle \sigma^x_N \rangle \sigma^x_1$,
we have $\mathcal{M}_j = -i J_x \sket{\sigma^x_1}$, and $\mathcal{E}_j = \mathcal{E}[\sigma^x_N]$
is the superoperator form of the expectation. 
Moreover ${\bf e_j}$ is the direction to the neighboring cluster involved.

\begin{figure}[!t]
  \centering
  \includegraphics[width=3.5in]{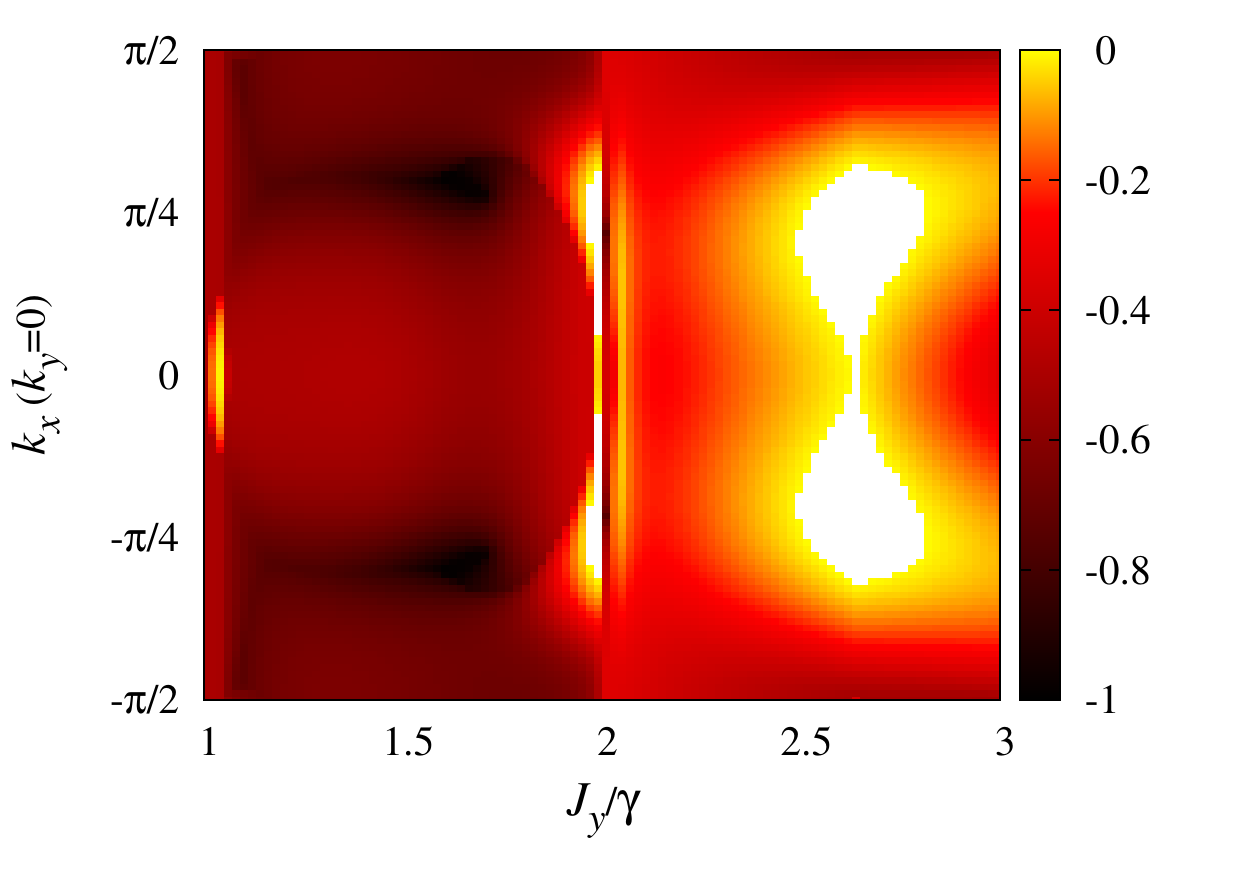}
  \includegraphics[width=3.5in]{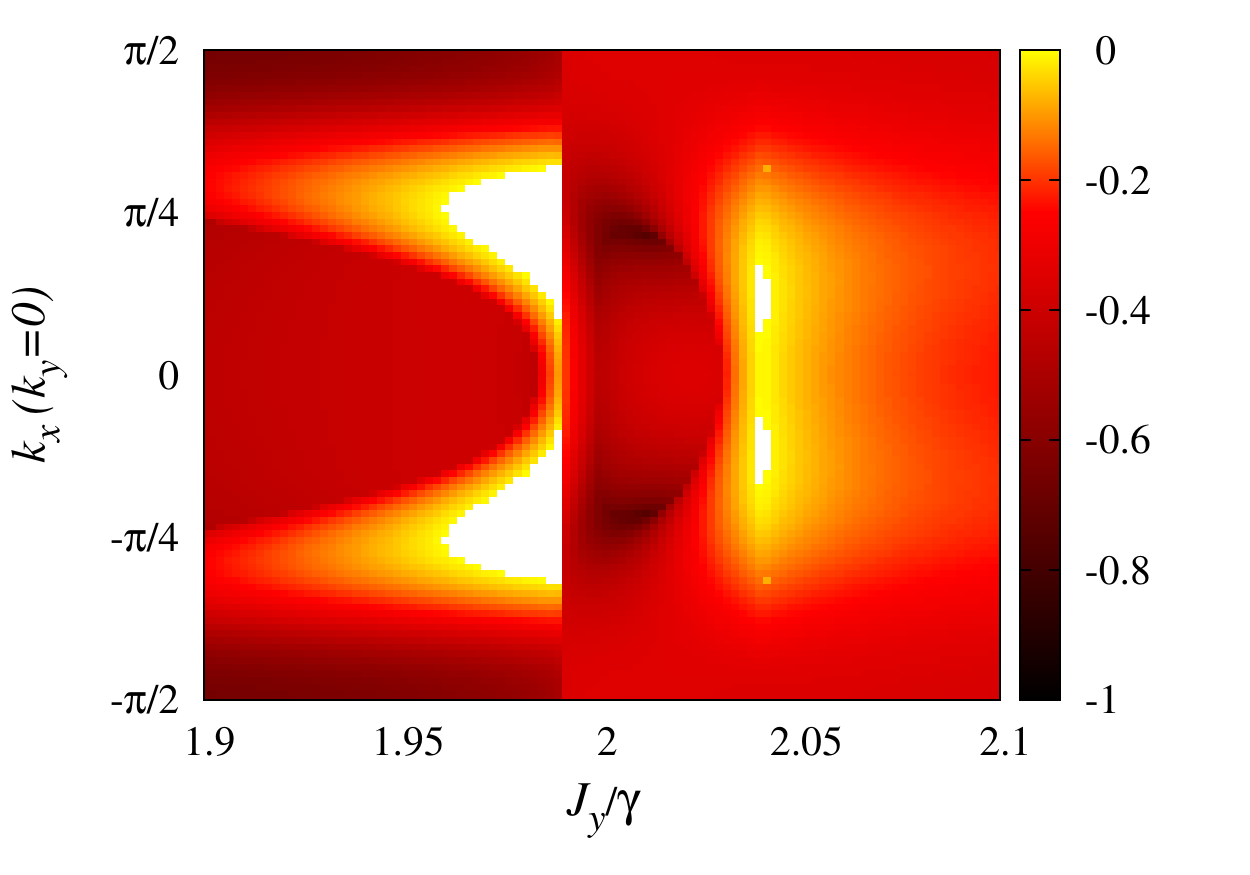}
  \caption{Upper panel: real part of most unstable eigenvalue (negative is stable) 
    as a function of $J_y$ and $k_x$ (for $k_y=0$). The parameters are set as in Fig.~\ref{fig:2D_Cut09}.
    Lower panel: high resolution plot with the same range of $J_y$ as in Fig.~\ref{fig:2x2_jump}, 
    corresponding to upward trace.}
  \label{fig:unstable_ev}
\end{figure}

When performing linear stability analysis, we expand the fluctuations in terms of plane waves 
\begin{equation}
  \sket{\rho_n} = \sket{\rho_0} + \sum_{\bf k} e^{i {\bf k} \cdot {\bf r_n}} \sket{\delta \rho_{\bf k}}
\end{equation}
so that the resulting equation of motion for $\sket{\delta \rho_{\bf k}}$ is
\begin{eqnarray} \nonumber
  \partial_t \sket{\delta\rho_{\bf k}} & = & 
  \bigg[ \mathcal{M}^0 + \sum_j \Big( \mathcal{E}_j \cdot \sket{\rho_{0}} \Big) \, \mathcal{M}_j \bigg] \sket{\delta \rho_{\bf k}} \\
  & & + \sum_j e^{i {\bf k} \cdot {\bf e_j}} \mathcal{M}_j \sket{\rho_0} \Big(\mathcal{E}_j \cdot \sket{\delta \rho_{\bf k}} \Big) \,.
\end{eqnarray}
The last term is a sum of rank-one matrices (since $\mathcal{M}_j \sket{\rho_0}$ is a vector, 
like $\mathcal{E}_j$). Thus we obtain
\begin{equation}
  \partial_t \sket{\delta\rho_{\bf k}} = \bigg[ \mathcal{M}^{0,\text{eff}} 
    + \sum_{\bf e} e^{i {\bf k} \cdot {\bf e}} \mathcal{M}^{1,{\bf e}} \bigg] \sket{\delta\rho_{\bf k}} \,,
  \label{eq:stab_matr}
\end{equation}
where in the second part, we have grouped terms with the same vector ${\bf e}$ together, 
as these all get the same ${\bf k}$ dependent factor. 
We then numerically compute the eigenvalues of the effective superoperator 
in Eq.~\eqref{eq:stab_matr} for each value of ${\bf k}=(k_x,k_y)$.
The most unstable eigenvalue is the one with the largest positive real part.
Since for a $\ell \times \ell$ cluster the vectors ${\bf e_j}$ must
be $\ell$ times the elementary lattice vectors, the range of lattice
momenta coming from the $\ell \times \ell$ cluster stability analysis is
restricted to the first Brillouin zone of the superlattice, $|k_j| < \pi/\ell$.

In Fig.~\ref{fig:unstable_ev} we plot the real part of the most unstable eigenvalue
as a function of the momentum $k_x$ and the coupling $J_y$ (for fixed $J_x = 0.9$).
We notice that the jump inside the FM phase occurs when there is an instability at finite $k$, 
around $|{\bf k}| = \pi/4$.
This suggests that the finite cluster size is responsible for the particular metamagnetic transition 
seen, and explains why the extent of the ordered phase reduces as larger clusters 
(capable of describing such short-range fluctuations) are used.
The transition to the normal state also occurs from a finite momentum instability, at small $|{\bf k}|$.
We also see an incommensurate finite-momentum instability at the rebirth of the FM phase,
for large $J_y$, thus signaling that probably the reappearance of the ordered phase is 
an artifact of the translationally invariant CMF Ansatz.
Finally we checked that the dispersion is almost isotropic in $k_x, k_y$.

\subsection{Three- and four-dimensional systems}
\label{sec:HigherD}

\begin{figure}[!t]
  \includegraphics[width=0.98\columnwidth]{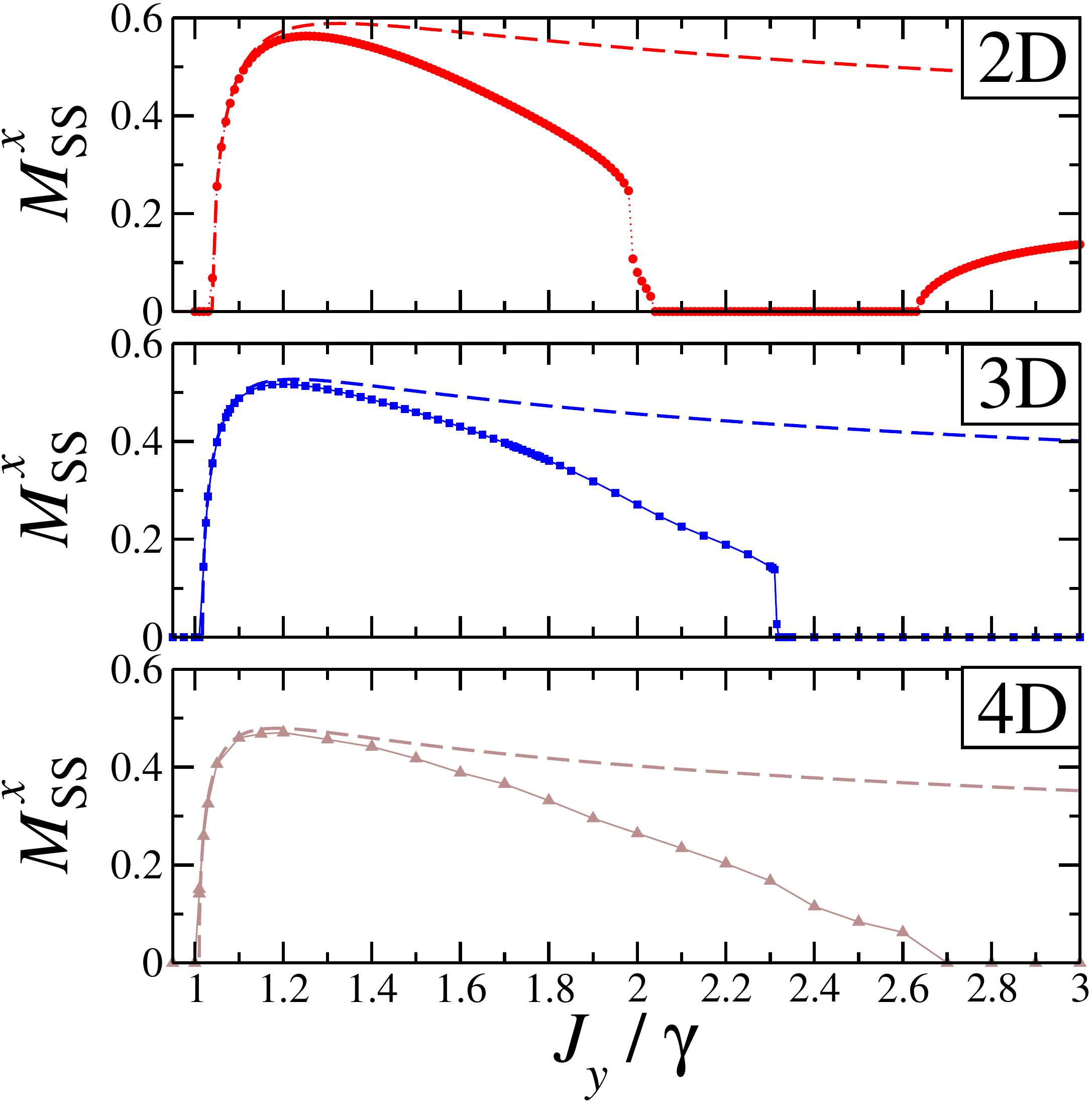}
  \caption{Cluster mean field analysis of the ferromagnetic order parameter 
    for a system defined on a square lattice geometry in different dimensionalities. 
    The upper panel refers to 2D ($z=4$), the middle one to 3D ($z=6$),
    the lower one to 4D ($z=8$).
    Dashed lines denote the MF predictions, while symbols and continuous curves
    are the results of numerical simulations, with clusters of size $L = 2^D$
    ($D$ being the system dimensionality).}
  \label{fig:HigherDim}
\end{figure}

For completeness we consider also the case of higher dimensions. 
Although not relevant for direct experimental realisations
it helps in completing the picture achieved so far; it may
also be possible to study (finite sized) four dimensional
systems by using synthetic dimensions as proposed recently
by Ozawa {\it et al.}~\cite{Ozawa}.
Mean-field results are expected to improve their validity with increasing 
the coordination number $z$ in the lattice.
It is therefore tempting to investigate systems in higher dimensionality 
by means of CMF techniques. Obviously, on increasing $d$, our ability in considering 
larger clusters goes drastically down. 
We checked the dependence on $d$ of the PM-FM transition by means of a mean-field analysis 
with clusters of size $L=2^d$. 
In these cases we looked again at the average on-site magnetisation along the $x$ axis. 
The results are displayed in Fig.~\ref{fig:HigherDim}. 

Naturally, the extent of the symmetry-broken phase region is increasing with the dimensionality, 
as it is apparent from Fig.~\ref{fig:HigherDim} (despite the value of the order parameter 
does not necessarily become larger). This supports the common wisdom of the validity 
of single-site mean field in high dimensions. What is surprising from Fig.~\ref{fig:HigherDim} 
is that even the four-dimensional system shows a critical value of $J_y$ beyond which 
the phase is paramagnetic. 
This is in sharp contrast with the mean-field result that does not capture 
this second critical point. Our limited analysis up to four-dimensions and for very small
clusters does not allow to draw conclusions in determining how/if the second critical point 
moves in higher dimensions. It is however an interesting point to be understood.

\subsection{Short-range correlations and Lindblad dynamics: Origin of the re-entrant paramagnetic phase}
\label{sec:physicalorig}

As discussed in detail in Sec~\ref{sec:2D}, our calculations show that, on improving the Ansatz 
for the steady-state density matrix by including short-range correlations, the critical points 
may shift from $J_y=+\infty$ to a finite value of $J_y$. This situation may appear as counterintuitive. 
It is indeed unlikely to occur at equilibrium, where the inclusion of short-range fluctuations 
may only lead to a shift of the boundary position of the order of the energy fluctuations 
inside the cluster [$O(z J_y)$ in this case]. 
In this section we want to explain the mechanism responsible for this behavior. 
This will also help us to elucidate the nature of the PM phase observed at large $J_y$ 
within the CMF approach and, consequently, the re-entrance to a disordered phase. 
To this aim it is sufficient to compare the single-site with the $2 \times 1$ (two-site) 
cluster cases. We consider this minimal cluster dimension for simplicity, 
since taking larger clusters would not add new ingredients to the understanding of the mechanism.

\begin{figure}[!t]
  \includegraphics[width=\columnwidth]{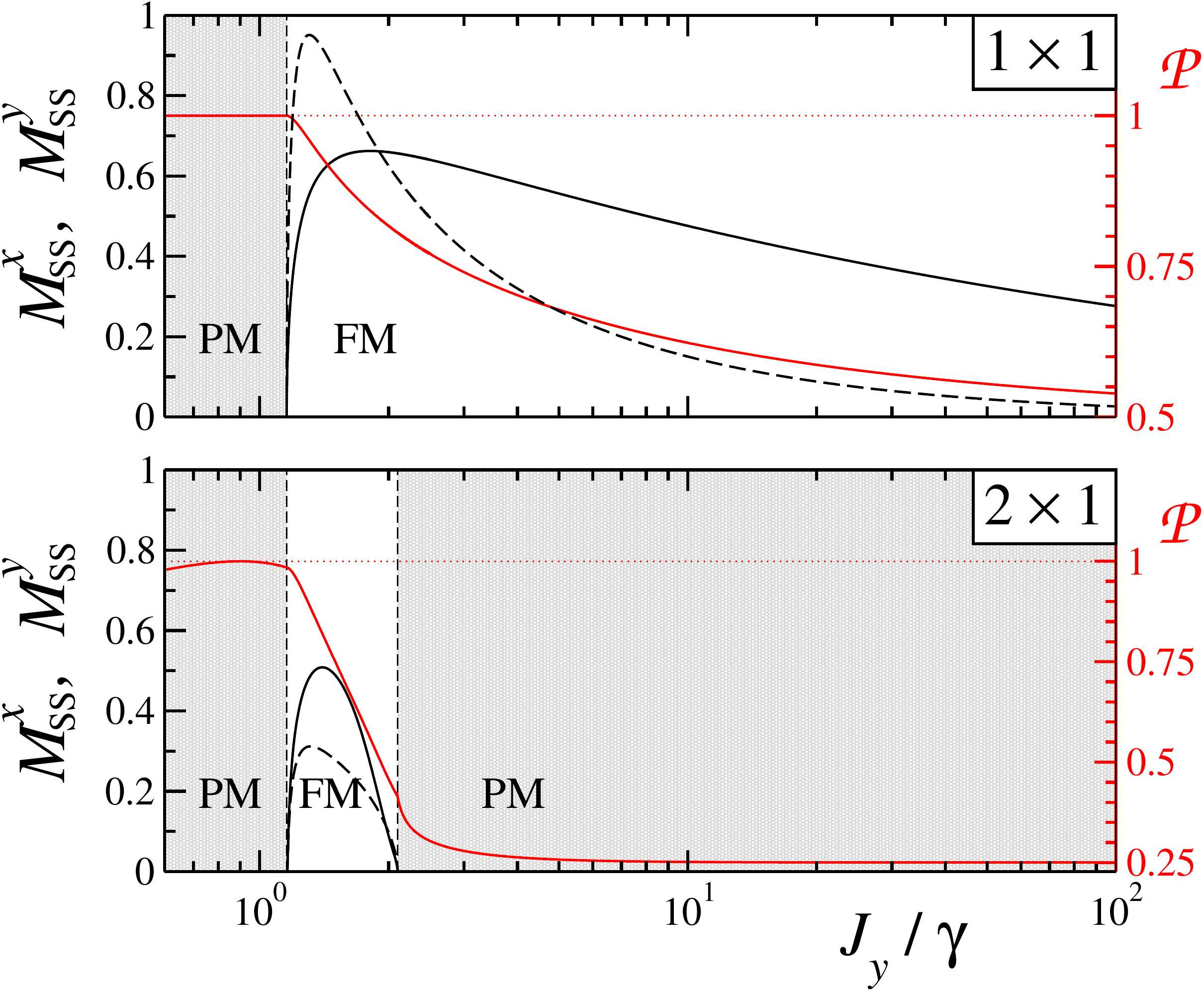}
  \caption{Magnetization along the $x$ (solid black line),
    $y$ (dashed black line) directions and purity 
    $\mathcal{P}={\rm Tr}[\rho_\mathcal{\scriptscriptstyle \rm SS}^2]$ (solid red line) 
    as a function of $J_y$, for fixed $J_x=0.9$, $J_z=1$.
    The single-site MF ($1 \times 1$, top panel) predicts a PM phase 
    for $J_y<J_y^c$ and a FM phase for $J_y>J_y^c$.
    On the contrary, according to a CMF analysis ($2 \times 1$, lower panel), 
    the FM phase shrinks down to a finite region 
    $J_y^c = J_y^{c {\rm (left)}} < J_y < J_y^{c {\rm (right)}}$ 
    allowing the re-entrance of a PM phase for $J_y > J_y^{c {\rm (right)}}$. }
  \label{fig:MagnMF_1vs2}
\end{figure}

First, it is important to stress that, already at the single-site MF, 
a steady state with vanishing spontaneous magnetization in all the directions 
is predicted in the limit $J_y \to +\infty$.
As shown in the top panel of Fig.~\ref{fig:MagnMF_1vs2}
(for fixed $J_x=0.9$ and $J_z=1$), two phases emerge: 
a PM for $J_y<J_y^c$ and a FM for $J_y>J_y^c$, with magnetization along $y$ (or equivalently 
along $x$) initially increasing, but then decreasing asymptotically toward zero as $J_y$ is increased.
This phenomenon is related to the progressive deterioration of the purity 
of the steady-state density matrix, $\mathcal{P}={\rm Tr}[\rho_{\scriptscriptstyle \rm SS}^2]$, 
for $J_y> J_y^c$.
This comes as a consequence of the out-of-equilibrium nature of the steady-state 
resulting from the interplay of driving and dissipation and cannot occur
at equilibrium, where an increasing coupling typically stabilizes the ordering.
A similar kind of behavior can be seen in a driven two-level system~\cite{Tian,Bishop} 
where increasing driving enhances the population, but suppresses the purity of the system, 
leading to a suppression of the homodyne ampltitude $|\langle \sigma^- \rangle |$ 
and of the purity when driven on resonance.
We see that $\mathcal{P}=1$ in the PM phase and then it  
decreases toward its minimal value ($\mathcal{P}=1/2$ in the case of a single-site cluster) 
as $J_y$ is increased beyond the critical value $J_y^{c}$.
This suggests the fact that the disordered phase detected for $J_y<J_y^{c}$ 
is different in nature compared to the one reached in the large-$J_y$ limit
for the cluster mean-field simulation.
The former is due to the stabilization of a fully polarized 
along the $z$ direction,
which coincides with the single-site MF solution for any $J_y < J_y^{c}$ 
(while it is the exact solution to the problem only for $J_x = J_y$).
The latter PM phase is a consequence of the fact that the steady-state 
for $J_y \to + \infty$ is fully mixed.

What is the effect of including short range correlations? 
In order to understand this point, let us consider more in detail the smallest cluster 
where this feature can be observed, namely a $2 \times 1$ plaquette.
As shown in the lower panel of Fig.~\ref{fig:MagnMF_1vs2}, the FM phase 
now shrinks to a finite region going from $J_y^{c {\rm (left)}}$ to $J_y^{c {\rm (right)}}$,
so that the PM for $J_y \to +\infty$ stabilizes over an extended region $J_y > J_y^{c {\rm (right)}}$.
The steady-state purity indicates a nearly pure state in the left PM region, 
$J_y < J_y^{c {\rm (left)}}$,
that has to be contrasted with a nearly fully mixed state in the right PM region,
$J_y > J_y^{c {\rm (right)}}$ (the minimal value for a two-site cluster is $\mathcal{P}=1/4$).
\red{
Thus, the exact inclusion of the nearest neighbor correlations allows for the reentrance 
of a PM phase for $J_y > J_y^{c {\rm (right)}}$. 
At single-site MF level, such PM phase appears only in the limiting case $J_y\to+\infty$ 
and then is never detectable for any finite value of the couplings.
We remark that the decreasing of the purity, and the consequent reduction of the magnetizations, 
as $J_y$ is increased is a common nonequilibrium feature of the two Ans\"atze ($1 \times 1$ and $2 \times 1$).
While in the $1 \times 1$ case the purity reduction is not enough to kill the FM order, 
in the $2 \times 1$ plaquette this reduction of purity is more prominent and the latter phenomenon (suppression of magnetization) occurs.
}

The equations of motion in the Heisenberg picture for magnetization
$\braket{\sigma^\beta_j}$ ($\beta=x,y,z$) are
\begin{eqnarray}
  \nonumber
  \partial_t \braket{\sigma^\beta_j} & = & 
  -2 \sum_{\alpha=x,y,z} J_\alpha \epsilon_{\alpha \beta \gamma} \left[ \braket{\sigma^\gamma_j} 
    \braket{\sigma^\alpha_{j+1}} + \braket{\sigma^\gamma_j \sigma^\alpha_{j+1}}\right] \\
  && \hspace*{1.5cm} -\frac\gamma2 \big[ \braket{\sigma^\beta_j} 
    + \delta_{\beta z}(\braket{\sigma^\beta_j}+2) \, \big],
  \label{2x1} 
\end{eqnarray}
where $\epsilon_{\alpha \beta \gamma}$ is the Levi-Civita symbol 
and $\delta_{\alpha\beta}$ is the Kronecker delta.
The steady-state density matrix in the $2 \times 1$ plaquette for $J_y>J_y^{c {\rm (right)}}$ 
can be analytically computed, and is {\it almost} fully mixed.
Therefore it can can be written as 
$\rho_{\scriptscriptstyle \rm SS}^{\scriptscriptstyle [2 \times 1]} 
\approx \rho^{\scriptscriptstyle [1]} \otimes \rho^{\scriptscriptstyle [2]}$. 
The two-point spin correlator appearing in Eq.~\eqref{2x1} can be thus decomposed as
\begin{equation}
  \label{spin_corr}
  \braket{\sigma^\gamma_j \sigma^\alpha_{j+1}} 
  = \braket{\sigma^\gamma_j} \braket{\sigma^\alpha_{j+1}}+ \braket{\Sigma^{\gamma,\alpha}_{j,j+1}},
\end{equation}
where $|\braket{\Sigma^{\gamma,\alpha}_{j,j+1}}| \ll 1$.
Inserting this expression into Eq.~\eqref{2x1} and exploiting translational invariance we get
\begin{equation}
  \label{2x1_bis}
  \partial_t \braket{\sigma^\beta_j} = \mathcal{L}^\beta_{\scriptscriptstyle [1\times1]} 
  -2 \, \sum_{\alpha} J_\alpha \epsilon_{\alpha\beta\gamma}\braket{\Sigma^{\gamma,\alpha}_{j,j+1}},
\end{equation}
where $\mathcal{L}^\beta_{\scriptscriptstyle [1\times1]}$ are the terms one would get from 
the single-site MF.
Equation~\eqref{2x1_bis} shows that spin-spin correlations correct the single-site 
MF equations of motion only through the small term $\braket{\Sigma^{\gamma,\alpha}_{j,j+1}}$.
On the other hand we know that, for $J_y > J_y^{c {\rm (right)}}$, 
the steady-state properties can change dramatically when considering a single site 
or a plaquette as a cluster: in the former case one gets a ferromagnet, 
while in the latter case one gets a paramagnet.
Spin-spin correlations, even if very weak, cannot be neglected and drastically modify 
the structure of the density matrix at long times. 
These conspire with the dynamically induced reduction of purity at large $J_y$, 
already visible for the single siste mean-field, to suppress the ordering altogether.
This is the key to understand 
the dramatic changes in the phase boundaries we presented in the previous sections.

We believe that the mechanism is generic and should be relevant for other driven-dissipative models as well.

\section{Conclusions}
\label{sec:concl}

In this work we introduced a cluster mean-field approach combined with quantum trajectories and 
tensor-network techniques to the study of the steady-state phase diagram in driven-dissipative systems. 
This approach allowed us to analyze the effect of short-range correlations. 
The result is somewhat unexpected. The whole structure of the phase diagram is radically modified 
in clear opposition to what typically happens in equilibrium phase transitions. 
In particular, we observed that the location of critical points may shift from 
infinite to finite values of the system parameters.
The reason underlying this behavior is related to the fact that, differently from equilibrium, 
spontaneous symmetry breaking is of pure dynamical nature and is not determined 
through a free-energy analysis. 
It is already known that in dissipative systems, energy-minimizing ferromagnetic phases 
may be destabilized, and replaced by incommensurate or antiferromagnetic order. 
Such behaviour has been extensively studied in classical
pattern forming systems~\cite{Cross1993}, including examples such as
active matter and flocking~\cite{vicsek, zia, Toner}.
As such, short range correlations can be expected to play a much greater role 
in dissipative than in equilibrium systems.
Accordingly, the topology of the phase diagram can significantly change. This appears clearly 
in Fig.~\ref{fig:diagram}, where the results from the single-site and the cluster mean-field 
analysis are compared. Furthermore, the cluster method hints at ordering with a non-trivial 
spatial pattern, a possibility which is not detected within the single-site mean-field Ansatz. 

The results that we highlighted here are amenable to an experimental verification. 
As discussed in Ref.~\cite{Lee2013}, the model considered in this paper can be implemented 
using trapped ions. Moreover, by changing external controls it is possible to explore 
the phase diagram, thus allowing to check the results of the present work.
Besides the examined system, we think that cluster approaches may be powerful 
in the general context of driven-dissipative systems, 
ranging from Rydberg atoms in optical lattices to cavity or opto-mechanical arrays.
Our findings point out the importance of the interplay between short-range fluctuations 
and dissipation in the physics emerging in such devices.

All the present analysis has been performed by considering a static mean field. 
It would be of great interest to extend these calculations so as to include also self-energy 
corrections as in the dynamical mean field, already extended to non-equilibrium for 
the single-site case~\cite{arrigoni2013}.

Finally we believe that a very interesting development, left for the future, 
is the determination of the non-Landau critical exponents. When successful, this will be 
an important step to establish the power of cluster techniques also in many-body open systems. 
On this perspective, the combination of our approach with the corner space renormalization method 
developed in Ref.~\cite{Finazzi2015} looks promising and some encouraging results 
have been already obtained~\cite{ciuti_private}.

\acknowledgments

We warmly thank useful discussions with I. Carusotto, C. Ciuti, M. Fleischhauer, S. Gopalakrishnan, M. Hafezi, and T. Lee. 
AB, LM, JK, RF, and DR acknowledge the Kavli Institute for Theoretical Physics, University of California, 
Santa Barbara (USA) for the hospitality and support during the completion of this work. 
This research was supported in part by the Italian MIUR via FIRB project RBFR12NLNA, by the EU integrated
projects SIQS, QUIC, by the National Science Foundation under Grant No.~NSF PHY11-25915, 
and by National Research Foundation, Prime Ministers Office, Singapore under its Competitive Research Programme
(CRP-QSYNC Award No.~NRF- CRP14-2014-02).
OV acknowledges support from the Spanish MINECO grant FIS2012-33152, the CAM research consortium QUITEMAD+, 
the U.S. Army Research Office through grant W911NF-14-1-0103, FPU MEC Grant and Residencia de Estudiantes.
JK acknowledges support from the EPSRC program ``TOPNES'' (EP/I031014/1).
JJ was supported by National Natural Science Foundation of China under Grants No.~11547119 and No.~11305021, by the Fundamental Research Funds for the Central Universities No.~DUT15RC(3)034, and by Natural Science Foundation of Liaoning Province No.~2015020110.
LM is supported by LabEX ENS-ICFP: ANR-10-LABX-0010/ANR-10-IDEX-0001-02 PSL*.


\begin{thebibliography}{100}
\bibitem{note01}
  In principle, it is even possible to proceed 
  as for the Hamiltonian case, and determine a self-consistent solution for the dark state 
  of the Liouvillean operator on the corresponding cluster (instead of the ground-state 
  of the cluster Hamiltonian).
  The computational effort is however huge, due to the size of the Liouvillian
  (e.g., in a system composed of $L$ spins, it is a $4^L \times 4^L$ non-Hermitian matrix).
\bibitem{note02}
  We also tried to break the symmetry by adding a small external magnetic field along the $x$ axis, 
  and then to extrapolate the value of the spontaneous magnetisation by calculating 
  the zero-field susceptibility. However the results we obtained are less clean than those found 
  by the analysis of the two-point correlations in Eq.~\eqref{eq:xx_Correl}.
\bibitem{note03}
  Note however that our code introduces a small perturbation to the state before each new parameter value 
  (to destabilise the normal state), thus the real bistability region will be larger than shown.
\end{thebibliography}
\end{document}